\documentclass{llncs}

\usepackage[german,english]{babel}
\usepackage{times,theorem}
\usepackage{latexsym,color,graphics}
\usepackage{url}
\usepackage{epsfig}
\usepackage{amssymb}
\usepackage{amsmath}
\usepackage{amsfonts}
\usepackage{diagrams11}
\diagramstyle[tight,centredisplay%
,dpi=600
,UglyObsolete
,heads=LaTeX%
]

\begingroup
\catcode`\~=11
\gdef\urltilde{\lower 0.6ex\hbox{~}}
\endgroup

\newcommand{\A}{\mathcal{A}} 
 \newcommand{\D}{\mathcal{D}}
\newcommand{\E}{\mathcal{E}} \newcommand{\F}{\mathcal{F}}
\newcommand{\G}{\mathcal{G}} 
\newcommand{\I}{\mathcal{I}} 
 \renewcommand{\L}{\mathcal{L}}
\newcommand{\M}{\mathcal{M}} \newcommand{\N}{\mathcal{N}}
 \renewcommand{\P}{\mathcal{P}}
 
\renewcommand{\S}{\mathcal{S}} \newcommand{\T}{\mathcal{T}}
 \newcommand{\V}{\mathcal{V}}
\newcommand{\W}{\mathcal{W}}

\title{Intensional RDB for  Big Data Interoperability}
\author{Zoran Majki\'c}
\authorrunning{Zoran Majki\'c}
\institute{ISRST, Tallahassee, FL, USA\\
\email{majk.1234@yahoo.com}\\ http://zoranmajkic.webs.com/}
\authorrunning{Zoran Majki\'c}

\newtheorem{propo}{Proposition}
\newtheorem{coro}{Corollary}

\newcount\pdfoutput
\begin{document}


\maketitle

\begin{abstract}
A new family of Intensional RDBs (IRDBs), introduced in  \cite{Majk14R},
extends the traditional RDBs with the Big Data and flexible and
'Open schema' features, able to preserve the user-defined relational
database schemas and all preexisting user's applications containing
the SQL statements for a deployment of such a relational data. The
standard RDB data is parsed into an internal vector key/value
relation, so that we obtain a column representation of data used in
Big Data applications, covering the key/value and column-based Big
Data applications as well, into a unifying RDB framework.
Such an IRDB architecture is adequate for the
massive migrations from the existing slow RDBMSs into this new
family of fast IRDBMSs  by offering a Big Data and new flexible
schema features as well. Here we present the interoperability features of the IRDBs
by permitting the queries also over the internal vector relations created by parsing
of each federated database in a given Multidatabase system.
We show that the SchemaLog
with the second-order syntax and ad hoc Logic Programming and its querying
fragment can be embedded into the standard SQL IRDBMSs, so that we obtain a full interoperabilty
features of IRDBs by using only the standard relational SQL
for querying both data and meta-data.
\end{abstract}


\section{Introduction}
Current RDBMSs
  were  obsolete and not ready to accept the new Big Data (BD) social-network Web applications in the
  last 10 years, so that the isolated groups of developers of these ad-hoc systems (e.g., Google, Amazon, LinkedIn, Facebook, etc..)
  could use only the ready old-known technics and development instruments in order
  to satisfy the highly urgent business market requirements.
 In an article of the
Computerworld magazine \cite{Com09a}, June 2009, dedicated to the
NoSQL meet-up in San Francisco is reported the following: "NoSQLers
came to share how they had overthrown the tyranny of slow, expensive
relational databases in favor of more efficient and cheaper ways of
managing data".   Moreover, the NoSQL movements advocate that
relational fit well  for data that is \emph{rigidly structured} with
relations and are
  designated for central deployments with single, large high-end machines, and not for distribution. Often
  they emphasize that SQL queries are expressed in a sophisticated
  language.
   \\
  However,  we can provide the BD infrastructure and physical level in a form of
   simpler structures, adequate to support the distributive and massive BigData query
   computations, by preserving the logically higher level interface to
   customer's applications. That is, it is possible to preserve the RDB interface to data, with
   SQL query languages  for the programmers of the software applications,
   with the "physical" parsing of data in more simple structures, able to deal with
   Big Data scalability in a high distributive computation framework. \\
   The first step to maintain the logical declarative (non-procedural) SQL query language
   level, is obtained by a revision of
   traditional RDBMSs is provided by developing H-store (M.I.T., Brown and Yale University), a next generation OLTP systems that operates on
   distributed clusters of shared-nothing machines where the data
   resides entirely in main memory, so that it was shown to
   significantly outperform (83 times) a traditional, disc-based
   DBMS. A more full-featured version of the system \cite{SADM08} that is able to
   execute across multiple machines within a local area cluster has
   been presented in August 2008. The data storage in H-store  is
   managed by a single-thread execution engine that resides
   underneath the transaction manager. Each individual site executes
   an autonomous instance of the storage engine with a fixed amount
   of memory allocated from its host machine. Multi-side nodes do
   not share any data structures with collocated sites, and hence
   there is no need to use concurrent data structures (every
   read-only table is replicated on all nodes nd other tables are
   divided horizontally into disjoint partitions  with a k-safety
   factor two).
   More recently, during 2010 and 2011, Stonebraker has been a critic of the NoSQL movement
   \cite{Ston10,SADM10}: "Here, we argue that using MR systems to perform
tasks that are best suited for DBMSs yields less than satisfactory
results \cite{SADM10a}, concluding that MR is more like an
extract-transform-load (ETL) system than a DBMS, as it quickly loads
and processes large amounts of data in an ad hoc manner. As such, it
complements DBMS technology rather than competes with it." After a
number of arguments about MR (MapReduction) w.r.t. SQL (with GROUP
BY operation), the authors conclude that parallel DBMSs provide the
same computing model as MR (popularized by Google and Hadoop to
process key/value data pairs), with the added benefit of using a
declarative SQL language. Thus, parallel DBMSs offer great
scalability over the range of nodes that customers desire, where all
parallel DBMSs operate (pipelining) by creating a query plan that is
distributed to the appropriate nodes at execution time. When one
operator in this plan send data to next (running on the same or a
different node), the data are pushed by the first to the second
operator (this concept is analog to the process described in my book
\cite{Majk14}, Section 5.2.1 dedicated to normalization of SQL terms
(completeness of the Action-relational-algebra category
\textbf{RA})), so that (differently from MR), the intermediate data
is never written to disk. The formal theoretical framework (the
database category \textbf{DB}) of the parallel DBMSs and the
semantics of database mappings between them is provided in Big Data
integration theory as
well \cite{Majk14}.\\
One step in advance in developing this NewSQL approach
\cite{Majk14R} is to extend the "classic" RDB systems with both
features:  to offer, on user's side, the standard RDB database
schema for SQL querying and, on computational side, the "vectorial"
relational database able to efficiently support the low-level
key/value data structures together, in the same logical SQL
framework. A new family of Intensional RDBs (IRDBs), introduced in
\cite{Majk14R}, which extends the traditional RDBs with the Big Data
and flexible and 'Open schema' features, able to preserve the
user-defined relational database schemas and all preexisting user's
applications containing the SQL statements for a deployment of such
a relational data. The standard RDB data is parsed into an internal
vector key/value relation, so that we obtain a column representation
of data used in Big Data applications, covering the key/value and
column-based Big Data applications as well, into a unifying RDB
framework.\\
The idea of having relational names as arguments goes back to
\cite{SmSm77} where the authors describe an algebraic operator
called SPECIFY that converts a single relation name into its
relation, but it is too far from our work by considering the Codd's
normal forms with their concept of "aggregation" and their concept
of "generalization" (as in Quillian's semantic networks) and trying
to mix both the database and AI areas.\\
The version of the relational model in which relation names may
appear as arguments of other relations is also provided in
\cite{Ross92}. In such an approach has been proposed an extension of
the relational calculus using HiLog as logical framework rather than
FOL, and they call this extension "relational HiLog". But is an
ad-hoc second-order logic where the variables may appear in
predicate names, and where there is ambiguity in programming
semantics because there is no distinction between variables,
function, constant and predicate symbols. Thus, it seams
syntactically to be a FOL but it has the particular semantics of the
Second Order Logic. In our approach we remain in the FOL syntax with
only new terms for the intensional elements (n-ary concepts) and
with only simple intensional extension of the standard Tarski's
semantics for the FOL. The extension of the relational algebra in
\cite{Ross92} is very different from our standard SQL algebra
framework: instead of this standard relational SQL algebra in
\cite{Ross92} are provided two relational algebra extensions:
"E-relational algebra" which extends standard relational algebra
with a set of expansion operators (these operators expand a set of
relation names into the union of their relations, with the
relational name as an extra argument, and hence not to a key/value
representations in Big Data used in our vector relation $r_V$); The
second is "T-relational algebra" which extends e-relational algebra
by a set of "totality" operators (to allow the access to the names
of all nonempty relations in the relational database). Thus, both
from the algebraical, structural  and logical framework this
approach is very different from our GAV Data integration model and a
minimal conservative intensional extension of the Tarski's FOL
semantics.\\
Another approaches  in which relation and attribute names may appear
as arguments of other relations are provided in the area of
integration of heterogeneous databases.  In \cite{LeBT92}   a simple
Prolog interpreter for a subset of F-logic was presented, but the
negation in Prolog is not standard as in FOL, and such an approach
is far from SQL models of data and querying of RDB databases. Also
in \cite{KrLK91} is  demonstrated  the power of using variables that
uniformly range over data and meta-data, for schema browsing and
interoperability, but their languages have a syntax closer to that
of logic programming languages, and far from that of SQL. The more
powerful framework (where the variables can range over the following
five sets: (i) names of databases in a federation; (ii) names of the
relations in a database; (iii) names of the attributes in the scheme
of a relations; (iv) tuples in a given relation in a database; and
(v) values appearing in a column corresponding to a given attribute
in a relation) is presented in SchemaSQL \cite{LaSS01} where the SQL
is extended in order to be able to query metadata, which in our case
is not necessary because we preserve the original RDB SQL in order
to be able to migrate from the RDB models into IRDB models with Big
Data vector relation without unnecessarily complications. In fact,
the extended relational algebra in SchemaSQL would be an non
desirable complication in order to obtain the flexible schema and
Big Data RDB features.\\
The \emph{interoperability} is the ability to share, interpret and
manipulate the information across the heterogeneous database systems
supported by Multidatabase systems (MDBS) in  a distributed network
by encompassing a heterogeneous mix of local database systems.
Languages based on higher-order logic have been used for the
interoperability by considering that the schematic information
should be considered as part of a database's information content.
The major advantage associated with such approaches, used in
SchemaLog \cite{LaSS97,LaSS01}, is the declaratively they derive
from their logical foundation. The weak points of the SchemaLOG is
that it uses the second-order logic syntax and an ad-hoc Prolog-like
fixpoint semantics. However, both of them are not necessary, as we will show
by using the IRDBs, just because the ordinary RDBs  have the FOL
syntax and do not need fixpoint semantics but ordinary FOL
semantics.\\
In what follows, we denote by $B^A$ the set of all functions from
$A$ to $B$, and by $A^n$ a n-folded cartesian product $A \times
...\times A$ for $n \geq 1$,  we denote by $\neg, \wedge, \vee,
\Rightarrow$ and $\Leftrightarrow$ the logical operators negation,
conjunction, disjunction, implication and equivalence, respectively.
For any two logical formulae $\phi$ and $\psi$ we define the XOR
logical operator $\underline{\vee}$ by  $\phi \underline{\vee} \psi$
logically equivalent to $(\phi \vee \psi) \wedge \neg (\phi \wedge
\psi)$.

\subsection{Syntax and semantics of SchemaLog \label{sec:Schema Log}}
 The SchemaLog is syntactically
higher-order clausal logic, and is based on the technical benefits
of soundness, completeness and compactness by a reduction to
first-order predicate calculus. It has a strictly higher expressive
power than first-order logic based on this syntax, differently from
IRDB which have the standard FOL
syntax but reacher intensional conservative Tarski's semantics.\\
The vocabulary of the SchemaLog language $\L_S$ consists of the
disjoint sets: $\G$ of k-ary ($k\geq 1$) functional symbols, $\S$ of
non-functional symbols (language constants, i.e., nullary functional
symbols), $\V$ of variables and usual logical connectives
$\neg,\vee,\wedge, \exists$ and $\forall$.\\
Every symbol in $\S$ and $\V$ is a term of the language, i.e., $\S
\bigcup \V \subseteq \T$. If $f \in \G$ is a n-ary function symbol
and $t_1,...,t_n$ are terms in $\T$ then
$f(t_1,...,t_n)$ is a term in $\T$. \\
An atomic formula of $\L_S$ is \emph{an expression} (note that it is
not a predicate-based atom of the FOL, that is, in SchemaLog we do not use the predicate letters) of the following forms \cite{LaSS97}:\\
 (i) $~~\langle db \rangle::\langle rel
\rangle[\langle tid \rangle: \langle attr \rangle
\rightarrow \langle val \rangle]$;\\
(ii) $~~\langle db \rangle::\langle rel
\rangle[\langle attr \rangle]$;\\
(iii) $~~\langle db \rangle::\langle rel \rangle$;\\
(iv) $~~\langle db \rangle$;\\
where $\langle db \rangle$ (the database symbols or names), $\langle
rel \rangle$ (the relational symbols or names), $\langle attr
\rangle$ (the attribute symbols or names), $\langle tid \rangle$
(the tuple-ids) and
$\langle val \rangle$  are the sorts  in $\S$ of $\L_S$.\\
The well-formed formulae (wff) of $\L_S$ are defined as usual: every
atom is a wff; $\neg \phi$, $\phi \vee \psi$, $\phi \wedge \psi$,
$(\exists x) \phi$ and $(\forall x)\phi$ are wffs of $\L_S$ whenever
$\phi$ and $\psi$ are wffs and $x \in \V$ is a variable.\\
A \emph{literal} is an atom or the negation of an atom. A
\emph{clause} is a formula of the form $\forall x_1,...,\forall
x_m(L_1 \vee ...\vee L_n)$ where each $L_i$ is a literal and
$x_1,...,x_m$ are the variables occurring in $L_1 \vee ...\vee L_n$.
A \emph{definite clause} is a clause in which one positive literal
is present and represented as $A \leftarrow B_1,...,B_n$ where $A$
is called the \emph{head} and $B_1,...,B_n$ is called the
\emph{body} of the definite-clause. A \emph{unit clause} is a clause
of the form  $A \leftarrow $, that is, a definite clause with an
empty body.\\
Let $\D$ be a nonempty set of elements (called "intensions"). A
\emph{semantic structure} \cite{LaSS97} of the language $\L_S$ is a
tuple $M = \langle
\D,\I,\I_{fun},\F \rangle$ where:\\
1. $\I:\S \rightarrow \D$ is a an interpretation of
non-function symbols in $\S$;\\
2. $\I_{fun}(f):\D^n \rightarrow \D$ is an interpretation of the
functional symbol $f \in \G$  of arity $n$;\\
3. $\F:\D \rightsquigarrow [\D \rightsquigarrow [\D \rightsquigarrow
[\D \rightsquigarrow \D]]]$, where $[A \rightsquigarrow B]$ denotes
the
set of all partial functions from $A$ to $B$.\\
To illustrate the role of $\F$, consider the atom $d:: r$. For this
atom to be true, $\F(\I(d))(\I(r))$ should be defined in $M$.
Similarly, for the atom $d::r[t:a \rightarrow v]$ to be true,
$\F(\I(d))(\I(r))\\(\I(a))(\I(t))$ should be defined in $M$ and
$\F(\I(d))(\I(r))(\I(a))(\I(t)) = \I(v)$.\\
A variable assignment $g$ is a function $g:\V \rightarrow \D$ (i.e.,
$g \in \D^{\V}$). We
extend it to all terms in $\T$ as follows:\\
$g(\overline{s}) = \I(\overline{s})$ for every $\overline{s} \in \S$;\\
$g(f(t_1,...,t_n)) = \I_{fun}(f)(g(t_1),...,g(t_n))$ where $f\in \G$
is a functional symbol of arity $n$ and $t_i$ are terms.\\
For a given set of terms $t_i \in \T$, $i = 1,2,...$ and the
formulae $\phi$ and $\psi$,  we define the satisfaction relation
$\models_g$ for a given assignment $g$ and the
structure $M$ as follows:\\
1. $M \models_g t_1~$ iff $~\F(g(t_1))$ is defined in $M$;\\
2.  $M \models_g t_1:: t_2 ~$ iff $~\F(g(t_1))(g(t_2))$ is defined in $M$;\\
3.  $M \models_g t_1:: t_2[t_3] ~$ iff $~\F(g(t_1))(g(t_2))(g(t_3))$ is defined in $M$;\\
4. $M \models_g t_1:: t_2[t_4:t_3 \rightarrow t_5] ~$ iff
$~\F(g(t_1))(g(t_2))(g(t_3))(g(t_4))$ is defined in $M$ and
$~\F(g(t_1))(g(t_2))(g(t_3))(g(t_4)) = g(t_5)$; \\
5. $M \models_g \phi \vee \psi ~$ iff $~M \models_g \phi$ or  $~M
\models_g \psi$;\\
6. $M \models_g \neg \phi~$ iff  not $~M \models_g \phi$;\\
7. $M \models_g (\exists x)\phi  ~$ iff for some valuation $g'$,
that may differ from $g$ only on the variable $x$, $~M \models_{g'}
\phi$.\\
$\square$\\
The specification of an extension of a RDB in this logic framework
can be done by specification of the Logic program with the (high)
number of unit and definite ground clauses (for each tuple in some
relational table of such an RDB) , which renders it unuseful for the
Big Data applications, because such a Logic Program would be
enormous. Moreover, we do not need to use the fixpoint semantics of
Logic programming for the definition of the extension of the RDBs
instead of the standard Tarski's semantics of the FOL. Thus, the
SchemaLog framework, defined for the Multidatabase interoperability,
can not be used for the interoperability in Big Data applications,
and hence we will show that SchemaLog can be reduced to intensional
RDB (IRDB) which
are designed for Big Data NewSQL applications.\\
 The plan of this paper is the following:\\
 In Section 2 we introduce the parsing of the RDBs into the Big Data vector
 relations, and then we present in Section 3 the intensional semantics for this new data
 structures, that is, of the intensional RDBs (IRDBs). The main
 Section 4 is dedicated to the Multidatabase IRDBs and we explain
 how a meta-data interoperability SchemaLog framework is embedded
 into the IRDBs Big Multidatabase systems.
\section{Vector databases of the IRDBs \label{sec:vector}}
 We will use the following RDB definitions, based on the
standard First-Order Logic (FOL) semantics:
\begin{itemize}
  \item  A \emph{database schema} is a pair $\A = (S_A , \Sigma_A)$ where $S_A$ is
  a countable set of relational symbols (predicates in FOL) $r\in \mathbb{R}$
  with finite arity
   $n = ar(r) \geq 1$ ($~ar:\mathbb{R} \rightarrow \mathcal{N}$),
disjoint from a countable infinite set $\textbf{att}$ of attributes
(a domain of $a\in \textbf{att}$ is a
 nonempty finite subset $dom(a)$ of a countable set of individual symbols
$\textbf{dom}$). For any $r\in \mathbb{R}$, the sort of $r$, denoted
by tuple $\textbf{a} = atr(r)= <atr_r(1),...,atr_r(n)>$ where all
$a_i = atr_r(m) \in \textbf{att}, 1\leq m \leq n$, must be distinct:
if we use two equal domains for different attributes then we denote
them by $a_i(1),...,a_i(k)$ ($a_i$ equals to $a_i(0)$). Each index
("column") $i$, $1\leq i \leq ar(r)$, has a distinct column name
$nr_r(i) \in SN$ where $SN$ is the set of names with $nr(r) =
<nr_r(1),...,nr_r(n)>$. A relation symbol $r \in \mathbb{R}$
represents the \emph{relational name} and can be used as an atom
$r(\textbf{x})$ of FOL with variables in $\textbf{x}$ assigned to
its columns, so that $\Sigma_A$ denotes a set of sentences (FOL
formulae without free variables) called \emph{integrity constraints}
\index{integrity constraints} of the sorted FOL with sorts in
$\textbf{att}$.
%
\item An \emph{instance-database} of a nonempty schema  $\A$ is given by
$A = (\A,I_T) = \{R =\|r\| = I_T(r) ~|~r \in S_A \}$ where $I_T$ is
a Tarski's FOL interpretation which satisfies \emph{all} integrity
constraints in $\Sigma_A$ and maps a relational symbol $r \in S_A$
into an n-ary relation $R=\|r\|\in A$. Thus, an instance-database
$A$ is a set of n-ary relations, managed by  relational database
systems.
\item We consider a rule-based
\emph{conjunctive query} over a database schema $\A$ as an
expression $ q(\textbf{x})\longleftarrow r_1(\textbf{u}_1), ...,
r_n(\textbf{u}_n)$, with finite $n\geq 0$, $r_i$ are the relational
symbols (at least one) in $\A$ or the built-in predicates (e.g.
$\leq, =,$ etc.), $q$ is a relational symbol not in $\A$ and
$\textbf{u}_i$ are free tuples (i.e., one may use either variables
or constants). Recall that if $\textbf{v} = (v_1,..,v_m)$ then
$r(\textbf{v})$ is a shorthand for $r(v_1,..,v_m)$. Finally, each
variable occurring in $\textbf{x}$ is a \emph{distinguished}
variable that must also occur at least once in
$\textbf{u}_1,...,\textbf{u}_n$. Rule-based conjunctive queries
(called rules) are composed of a subexpression $r_1(\textbf{u}_1),
...., r_n(\textbf{u}_n)$ that is the \emph{body}, and
 the \emph{head} of this rule $q(\textbf{x})$.\\
 The deduced head-facts  of a conjunctive query $q(\textbf{x})$ defined over an instance $A$ (for a given Tarski's
 interpretation $I_T$ of  schema $\A$) are
 equal to $\|q(x_1,...,x_k)\|_A = \{<v_1,...,v_k> \in \textbf{dom}^k ~|~ \exists \textbf{y}(r_1(\textbf{u}_1)\wedge
 ...\wedge
r_n(\textbf{u}_n))[x_i/v_i]_{1\leq i \leq k}$ is true in $A \}
 = I_T^*(\exists \textbf{y}(r_1(\textbf{u}_1)\wedge ...\wedge
r_n(\textbf{u}_n)))$, where the $\textbf{y}$ is a set of variables
which are not in the head of query, and $I_T^*$ is the unique
extension of $I_T$ to all formulae.
Each conjunctive query corresponds
to a "select-project-join" term $t(\textbf{x})$
 of SPRJU algebra obtained from the formula $\exists \textbf{y}(r_1(\textbf{u}_1)\wedge ...\wedge
r_n(\textbf{u}_n))$.
 \item
  We consider a finitary \emph{view} as a union of a finite set $S$ of conjunctive  queries with the same
 head $q(\textbf{x})$ over a schema $\A$, and from the equivalent
algebraic point of view, it is a "select-project-join-rename +
union" (SPJRU) finite-length term $t(\textbf{x})$ which corresponds
to union of the terms of conjunctive queries in $S$.
 A
materialized view of an instance-database $A$ is an n-ary relation
$R = \bigcup_{q(\textbf{x}) \in S}\|q(\textbf{x})\|_A$.
 \end{itemize}
 Recall that two relations $r_1$ and $r_2$ are union-compatible iff
$\{atr(r_1)\} = \{atr(r_2)\}$. If a relation $r_2$ is obtained from
a given relation $r_1$ by permutating its columns, then we tell that
they are not equal (in set theoretic sense) but that they are
equivalent. Notice that in the RDB theory the two equivalent
relations are considered equal as well. In what follows, given any
two lists (tuples), $\textbf{d} = <d_1,...,d_k>$ and $\textbf{b} =
<b_1,...,b_m>$ their concatenation $<d_1,...,d_k,b_1,...,b_m>$ is
denoted by $\textbf{d} \& \textbf{b}$, where $'\&'$ is the symbol
for concatenation of the lists. \\
The set of basic relation algebra operators are:\\
1. Rename is a unary operation written as $\_~$ RENAME
$~name_1~$AS$~name_2 $
  where the result is identical to input argument (relation) $r$ except that the
  column $i$ with  name $nr_r(i) = name_1$ in all tuples is renamed to $nr_r(i) =
  name_2$.\\
2. Cartesian product is a binary operation $\_~ $TIMES $\_~$,
written also as $\_~ \bigotimes \_~$,
   such that for the relations $r_1$ and $r_2$, first we do the rename normalization
   of $r_2$ (w.r.t. $r_1$).\\
3. Projection is a unary operation written as $\_~[S]$, where
  $S$ is a tuple of column names such that for a relation $r_1$
  and $S = <nr_{r_1}(i_1),...,nr_{r_1}(i_k)>$, with $k \geq 1$ and
  $1 \leq i_m \leq ar(r_1)$ for $1 \leq m \leq k$, and $i_m \neq i_j$ if $m \neq j$, we define the relation $r$ by:
$~r_1[S]$,\\
with $\|r\| =  \|r_1\|$ if $\exists name \in S.name \notin nr(r_1)$;
otherwise $\|r\| =\pi_{<i_1,...,i_k>}(\|r_1\|)$, where $nr_r(m) =
nr_{r_1}(i_m)$, $atr_r(m) = atr_{r_1}(i_m)$, for $1 \leq m \leq
k$.\\
4. Selection is a unary operation written as $\_~ $WHERE$ ~C$,
where  a condition $C$ is a finite-length logical formula that
consists of atoms
  $'(name_i ~\theta ~name_j)'~$ or $~'(name_i ~\theta ~\overline{d})~'$,
  with built-in predicates $\theta \in \Sigma_\theta \supseteq \{ \doteq,>,< \}$,  a constant $\overline{d}'$,
    and the logical operators  $\wedge$ (AND),  $\vee$ (OR) and  $\neg$
  (NOT), such that for a relation $r_1$ and $name_i$, $name_j$ the names of its columns, we define the relation $r$ by\\
  $ r_1  ~$WHERE$~ C$,\\
  as the relation with $atr(r) = atr(r_1)$ and the function $nr_r$ equal to $nr_{r_1}$, where $\|r\|$ is composed
  by the tuples in  $\|r_1\|$ for which $C$ is satisfied.\\
5. Union  is a binary operation written as $\_~$ UNION $\_~$,  such
that for two
    union-compatible relations $r_1 $ and $r_2$, we define the relation $r$ by:
$r_1~$ UNION $~r_2$,\\
where $\|r\| \triangleq \|r_1\| \bigcup \|r_2\|$, with $atr(r) =
atr(r_1)$, and the functions $atr_r = atr_{r_1}$, and $nr_r =
nr_{r_1}$.\\
6. Set difference  is a binary operation written as $\_~ $MINUS$
\_~$  such that for two
  union-compatible relations $r_1$ and $r_2$, we define the relation $r$ by:
$r_1 ~$MINUS$~ r_2$,\\
where $\|r\|  \triangleq \{\textbf{t}~|~\textbf{t}\in \|r_1\|$ such
that $\textbf{t}\notin \|r_2\|\}$, with $atr(r) = atr(r_1)$, and the
functions $atr_r = atr_{r_1}$, and
$nr_r = nr_{r_1}$.\\
Natural join $\bowtie_S$ is a binary operator, written as $(r_1
\bowtie_S r_2)$, where $r_1$ and $r_2$ are the relations. The result
of the natural join is the set of all combinations of tuples in
$r_1$ and $r_2$ that are equal on their common attribute names. In
fact, $(r_1 \bowtie_S r_2)$ can be obtained by creating the
Cartesian product $r_1\bigotimes r_2$ and then by execution of the
Selection with the condition $C$ defined as a conjunction of atomic
formulae $(nr_{r_1}(i) = nr_{r_2}(j))$ with
$(nr_{r_1}(i),nr_{r_2}(j)) \in S$ (where $i$ and $j$ are the columns
of the same attribute in $r_1$ and $r_2$, respectively, i.e.,
satisfying $atr_{r_1}(i) = atr_{r_2}(j)$) that represents the
equality of the common attribute names of $r_1$ and $r_2$.\\
 We are able to define a new
relation with a single tuple $\langle
\overline{d}_1,..,\overline{d}_k \rangle, k \geq 1$ with the given
list of attributes $\langle a_1,..,a_k \rangle$, by the following
finite length expression,\\
EXTEND (...(EXTEND $r_\emptyset$ ADD $a_1,name_1$
AS$\overline{d}_1)...)$ ADD $a_k,name_k$ AS $\overline{d}_k$, or
equivalently by
$r_\emptyset \langle a_1,name_1, \overline{d}_1\rangle  \bigotimes...\bigotimes r_\emptyset\langle a_k,name_k, \overline{d}_k\rangle$,\\
 where $r_\emptyset$ is the empty
type relation  with $\|r_\emptyset \|= \{<>\}$, $ar(r_\emptyset) =
0$ introduced in Definition \ref{def:bealer}, and empty functions
$atr_{r_\emptyset}$ and $nr_{r_\emptyset}$. Such single tuple
relations can be used for an insertion in a given relation (with the
same list of attributes) in
what follows.\\
 \textbf{Update operators.} The three update operators,
'UPDATE', 'DELETE' and 'INSERT'  of the Relational algebra, are
derived operators from these previously defined operators  in the
following way:
\begin{enumerate}
  \item Each algebraic formulae 'DELETE FROM $~ r$ WHERE C' is
equivalent to the formula '$r$ MINUS ($r$ WHERE $C$)'.
  \item Each algebraic expression (a term) 'INSERT INTO $~r[S]$ VALUES (list of values)', 'INSERT INTO $~r[S]$ AS SELECT...',
is equivalent to  '$r~$ UNION $~r_1$' where the union compatible
relation $r_1$ is a one-tuple relation
(defined by list) in the first,  or a relation defined by 'SELECT...' in the second case.\\
  \item Each algebraic expression 'UPDATE $~r$ SET $[nr_r(i_1)= e_{i_1} ,..., nr_r(i_k) =
e_{i_k}]$ WHERE  $C$', for  $n =ar(r)$, where $e_{i_m}, 1 \leq i_m
\leq n$ for $1 \leq m\leq k$ are the expressions and $C$ is a
condition, is equal to the formula '($r$
WHERE  $\neg C$) UNION $r_1$' , where $r_1$ is a relation expressed by\\
  (EXTEND(...(EXTEND ($r$ WHERE $C$) ADD $att_r(1),name_1$ AS $e_1$)...) ADD $att_r(n),name_n$ AS
 $e_{n})[S]$,\\
 such that for each $1 \leq m \leq n$, if $m \notin \{i_1,...,i_k \}$ then $e_m  =
 nr_r(m)$, and $S = <name_1,...,name_n>$.
\end{enumerate}
Consequently, all update operators of the relational algebra can be
obtained by addition of these 'EXTEND $\_~$ ADD $a,name$ AS $e$'
operations. \\
Let us define the $\Sigma_R$-algebras sa follows (\cite{Majk14},
Definition 31 in Section 5.1):
\begin{definition} \label{def:relAlg}
  We denote the algebra of the set of operations, introduced previously in
this section (points from 1 to 6 and \emph{EXTEND} $\_~$ \emph{ADD}
$a,name$ \emph{AS} $e$) with additional nullary operator
(empty-relation constant) $\perp$, by $\Sigma_{RE}$. Its subalgebra
without $\_~ $\emph{MINUS}$ \_~$ operator is denoted by
$\Sigma_R^+$, and without $\perp$ and unary operators \emph{EXTEND}
$\_~$ \emph{ADD} $a,name$ \emph{AS} $e$  is denoted by $\Sigma_R$
(it is the "select-project-join-rename+union" (\emph{SPJRU})
subalgebra). We define the  set of terms $\T_PX$ with variables in
$X$
of this $\Sigma_R$-algebra (and analogously for the terms $\T_P^+X$
of $\Sigma_R^+$-algebra),
inductively as follows:\\
1.  Each relational symbol (a variable) $r \in X \subseteq
\mathbb{R}$ and  a constant (i.e., a nullary operation) is a term in
 $\T_P X$;\\
2. Given any term $t_R \in \T_P X$ and an unary operation $o_i \in
\Sigma_R$, $o_i(t_R)\in  \T_P X$;\\
3. Given any two terms $t_R, t'_R \in \T_P X$ and a binary operation
$o_i \in \Sigma_R$, $o_i(t_R,t'_R) \in \T_P X$.\\
We define the evaluation of terms in $\T_P X$, for $X = \mathbb{R}$,
by extending the assignment  $\|\_~\|:\mathbb{R} \rightarrow
\underline{\Upsilon}$, which assigns a relation to each relational
symbol (a variable) to all terms by the function $\|\_~\|_{\#}:\T_P
\mathbb{R} \rightarrow \underline{\Upsilon}$ (with $\|r\|_{\#} =
\|r\|$), where $\underline{\Upsilon}$ is the universal database
instance (set of all relations for a given universe $\D$). For a
given term $t_R$ with relational symbols $r_1,..,r_k \in
\mathbb{R}$, $\|t_R\|_{\#}$ is the relational table obtained from
this expression for the given set of relations $\|r_1\|,...,\|r_k\|
\in \underline{\Upsilon}$, with the constraint that \\ $\|t_R $
\emph{UNION} $t'_R\|_{\#} = \|t_R \|_{\#} \bigcup \|t'_R\|_{\#}$ if
the relations $\|t_R \|_{\#}$ and $\|t'_R \|_{\#}$ are
union compatible; $\perp = \{<>\} = \|r_\emptyset\|$ (empty relation) otherwise.\\
We say that two terms $t_R, t'_R \in \T_P X$ are equivalent (or
equal), denoted by $t_R \approx t'_R$, if for all assignments
$~\|t_R\|_{\#} = \|t'_R\|_{\#}$.
\end{definition}
The principal idea for the IRDBs (intensional RDBs) introduced in
\cite{Majk14R} is to use an analogy with a GAV Data Integration
\cite{Lenz02,Majk14} by using the database schema $\A = (S_A,
\Sigma_A)$ as a  global relational schema, used as a
user/application-program interface for the query definitions in SQL,
and to represent the source database of this Data Integration system
by  parsing of the RDB instance $A$ of the schema $\A$ into a single
vector relation $\overrightarrow{A}$. Thus, the original SQL query
$q(\textbf{x})$ has to be equivalently rewritten over
(materialized) source vector database $\overrightarrow{A}$.\\
 In fact, each $i$-th column value  $d_i$ in a tuple $\textbf{d} = \langle d_1,...,d_i,...,d_{ar(r)} \rangle$
   of a relation $R_k = \|r_k\|, r_k \in S_A$,  of the
instance database $A$ is determined by the free dimensional
coordinates: relational name $nr(r)$, the attribute name $nr_r(i)$
of the i-th column, and the tuple index $Hash(\textbf{d})$ obtained
by hashing the string of the tuple $\textbf{d}$. Thus, the
relational schema of the vector relation is composed by the four
attributes, relational name, tuple-index, attribute name, and value,
i.e., \verb"r-name",   \verb"t-index", \verb"a-name" and
\verb"value", respectively, so that if we assume $r_V$ (the name of
the database $\A$) for the name of this vector relation
$\overrightarrow{A}$ then this
relation can be expressed by the quadruple\\
$r_V$(\verb"r-name",   \verb"t-index", \verb"a-name",
\verb"value"),\\
and the parsing of any RDB instance $A$ of a schema $\A$ can be
defined as:
\begin{definition} \textsc{Parsing RDB instances:}
\label{def:parsing} \\Given a database instance $A =
\{R_1,...,R_n\}$, $n\geq 1$, of a RDB schema $\A = (S_A,\Sigma_A)$
with $S_A = \{r_1,...,r_n\}$ such that $R_k = \|r_k\|, k = 1,...,n$,
then the extension $\overrightarrow{A} = \|r_V\|$ of the vector
relational symbol (name) $r_V$ with the schema $r_V$(\verb"r-name",
  \verb"t-index", \verb"a-name",
\verb"value"),  and \emph{NOT NULL} constraints for all its four
attributes,
and with the primary key composed by the first three attributes, is defined by:\\
we define the  operation \emph{\underline{PARSE}} for  a tuple
$\textbf{d} =\langle d_1,...,d_{ar(r_k)}\rangle$ of the relation
$r_k \in S_A$ by the
mapping\\
 $(r_k,\textbf{d}) ~~\mapsto ~~\{\langle r_k,
 Hash(\textbf{d}),nr_{r_k}(i),d_i \rangle |~  d_i \emph{NOT NULL},
1\leq i \leq ar(r_k)\}$,
so that\\
 (1) $~~~\overrightarrow{A} = \bigcup_{r_k \in S_A,\textbf{d} \in
 \|r_k\|} \emph{\underline{PARSE}}(r_k,\textbf{d})$.\\
 Based on the vector database representation
 $\|r_V\|$ we define a GAV Data Integration system  $\I = \langle \A, \S,
\M \rangle$ with the global schema $\A = (S_A, \Sigma_A)$, the
source schema $\S = (\{r_V\},\emptyset)$, and the set of mappings
$\M$ expressed by the tgds (tuple generating
dependencies)\\
(2) $~~~\forall y,x_1,...,x_{ar(r_k)}(((r_V(r_k,y,nr_{r_k}(1),x_1)
~\underline{\vee}~ x_1 \emph{NULL})\wedge ...\\...\wedge
(r_V(r_k,y,nr_{r_k}(ar(r_k)),x_{ar(r_k)})~\underline{\vee}~
x_{ar(r_k)} \emph{NULL})) \Rightarrow r_k(x_1,...,x_{ar(r_k)}))$,\\
for each $r_k \in S_A$.
\end{definition}
The operation \underline{PARSE} corresponds to the parsing of the
tuple $\textbf{v}$ of the relation $r_k \in S_A$ of the user-defined
database schema $\A$ into a number of tuples of the vector relation
$r_V$. In fact, we can use this operation for virtual
inserting/deleting of the tuples in the user defined schema $\A$,
and  store them only in the vector relation $r_V$. This operation
avoids to materialize the user-defined (global) schema, but only the
source database $\S$, so that each user-defined SQL query has to be
equivalently rewritten over the source database (i.e., the big table
$\overrightarrow{A} = \|r_V\|$) as in standard
FOL Data Integration systems.\\
 Notice that
this parsing defines a kind of GAV Data Integration systems, where
the source database $\S$ is composed by the unique
vector relation
$\|r_V\| = \overrightarrow{A}$ (Big Data) which does not contain
NULL values, so that we do not unnecessarily save the NULL values of
the user-defined relational tables $r_k \in S_A$ in the main
memories of the parallel RDBMS used to horizontal partitioning of
the unique big-table $\overrightarrow{A}$. Moreover, any adding of
the new columns to the user-defined schema $\A$ does not change the
table $\overrightarrow{A}$, while the deleting of a $i$-th column of
a relation $r$ will delete all tuples $r_V(x,y,z,v)$ where $x =
nr(r)$ and $z = nr_r(i)$ in the main memory of the parallel RDBMS.
Thus, we obtain very \emph{schema-flexible} RDB model for Big
Data.\\
The intensional Data Integration system $\I = (\A,\S,\M)$ in
Definition \ref{def:parsing} is used in the way that the global
schema is only virtual (empty) database with a user-defined schema
$\A = (S_A, \Sigma_A)$ used to define the SQL user-defined query
which then has to be equivalently rewritten over the vector relation
$r_V$ in order to obtain the answer to this query. Thus, the
information of the database is stored only in the big table
$\|r_V\|$. Thus, the materialization of the original user-defined
schema $\A$ can be obtained by the following operation:
\begin{definition} \textsc{Materialization of the RDB}
\label{def:matz} \\Given a user-defined RDB schema $\A = (S_A,
\Sigma_A)$ with $S_A = \{r_1,...,r_n\}$ and a big vector table
$\|r_V\|$, the non SQL operation \emph{\underline{MATTER}} which
materializes the schema $\A$ into its instance database $A =
\{R_1,...,R_n\}$ where $R_k = \|r_k\|$, for $k = 1,...,n$, is given
by the following mapping, for
any  $R \subseteq \|r_V\|$:\\
$(r_k, R) ~~~\mapsto ~~~ \{\langle v_1,...,v_{ar(r_k)}\rangle ~|~
\exists y \in \pi_2(R) ((r_V(r_k,y,nr_{r_k}(1),v_1)
~\underline{\vee}~ v_1 \emph{NULL})\wedge ...\\...\wedge
(r_V(r_k,y,nr_{r_k}(ar(r_k)),v_{ar(r_k)})~\underline{\vee}~
v_{ar(r_k)} \emph{NULL})) \}$,\\
 so that the materialization of the schema $\A$ is defined by\\
   $R_k =\|r_k\| \triangleq \emph{\underline{MATTER}}(r_k, \|r_V\|)$ for each $r_k \in S_A$.
\end{definition}
The canonical models of the intensional Data Integration system $\I
= (\A,\S,\M)$ in Definition \ref{def:parsing} are the instances $A$
of the schema $\A$ such that\\
 $\|r_k\| =$ \underline{MATTER}$(r_k,
\bigcup_{\textbf{v}\in \|r_k\|}$
\underline{PARSE}$(r_k,\textbf{v}))$, that is,
when\\
$A = \{$\underline{MATTER}$(r_k, \overrightarrow{A}) ~|~r_k \in S_A\}$.\\
We say that an extension $\|t_R\|_{\#}$,  of a term $t_R \in \T_PX$,
is \emph{vector relation} of the \emph{vector view} denoted by
$\overrightarrow{t_R}$ if the type of $\|t_R\|_{\#}$ is equal to the
type of the vector relation $r_V$.\\
Let $R = \|\overrightarrow{t_R}\|_{\#}$ be the relational table with
the four attributes (as $r_V$)
\verb"r-name",\verb"t-index"\\\verb"a-name" and \verb"value", then
 its used-defined view representation can be derived as follows:
\begin{definition} \textsc{View Materialization}: \label{def:view-mater}
Let $t_R \in \T_PX$ be a user-defined SPJU
(Select-Project-Join-Union) view over a database schema $\A =
(S_A,\Sigma_A)$ with the type (the tuple of the view columns)
$\mathfrak{S} =
\langle(r_{k_1},name_{k_1}),...,(r_{k_m},name_{k_m})\rangle$, where
the $i$-th column $(r_{k_i},name_{k_i})$ is the column  with name
equal to $name_{k_i}$ of the relation name $r_{k_i}\in S_A$, $1\leq
i \leq m$, and $\overrightarrow{t_R}$ be the rewritten query over
$r_V$. Let $R = \|\overrightarrow{t_R}\|_{\#}$ be the resulting
relational table with the four attributes (as $r_V$)
\verb"r-name",\verb"t-index",\verb"a-name" and \verb"value". We
define the operation \emph{\underline{VIEW}} of the transformation
of $R$ into the user defined view
representation by:\\
$\emph{\underline{VIEW}}(\mathfrak{S},R) = \{\langle
d_1,...,d_m\rangle ~|~\exists ID \in \pi_3(R), \forall_{1\leq i \leq
m}(\langle r_{k_i},ID,name_{k_i},d_i \rangle \in R)$; otherwise set
$d_i$ to \emph{NULL} $\}$.
\end{definition}
Notice that we have  $\|r_k\| =$ \underline{VIEW}$(\mathfrak{S},R) =
$\underline{MATTER}$(r_k,R)$ for each $r_k \in S_A$ with $R =
\bigcup_{\textbf{d} \in\|r_k\|}$\underline{PARSE}$(r_k,\textbf{d})$,
and $\mathfrak{S} =
\langle(r_{k},nr_{r_k}(1)),...,(r_{k},nr_{r_k}(ar(r_k)))\rangle$,
and hence the nonSQL operation \underline{MATTER} is a special case
of the operation \underline{VIEW}.\\
 For any original user-defined query (term)
$t_R$ over a user-defined database schema $\A$, by
$\overrightarrow{t_R}$ we denote the equivalent (rewritten) query
over the vector relation $r_V$. We have the following important
result for the IRDBs:
\begin{propo} \label{prop:newSQL} There exists a complete algorithm for the
 term rewriting of any user-defined SQL term $t_R$ over a schema $\A$, of the
full  relational algebra $\Sigma_{RE}$ in Definition
\ref{def:relAlg},  into an equivalent vector query
$\overrightarrow{t_R}$ over the vector relation $r_V$. \\If $~t_R$
is a SPJU  term (in Definition \ref{def:view-mater}) of the type
$\mathfrak{S}$ then $\|t_R\|_{\#} =
\emph{\underline{VIEW}}(\mathfrak{S},\|\overrightarrow{t_R}\|_{\#})$.
\end{propo}
The proof can be find in \cite{Majk14R}. This proposition
demonstrates that the IRDB is full SQL database, so that each
user-defined query over the used-defined RDB database schema $\A$
can be equivalently transformed by query-rewriting into a query over
the vector relation $r_V$. However, in the IRDBMSs we can use more
powerful and efficient algorithms in order to execute
each original user-defined query over the vector relation $r_V$.\\
 Notice that this
proposition demonstrates that the IRDB is a kind of GAV Data
Integration System $\I = (\A,\S,\M)$ in Definition \ref{def:parsing}
where we do not materialize the user-defined schema $\A$ but only
the vector relation $r_V \in \S$ and each original query
$q(\textbf{x})$ over the empty schema $\A$ will be rewritten into a
vector query $\overrightarrow{q(\textbf{x})}$ of the type
$\mathfrak{S}$ over the vector relation $r_V$, and then the
resulting view
\underline{VIEW}$(\mathfrak{S},\|\overrightarrow{q(\textbf{x})}\|_{\#})$
will be returned to user's application. The operators
\underline{PARSE}, \underline{MATTER} and \underline{VIEW} can be
represented \cite{CuGG04} as derived algebraic operators of the
(UN)PIVOT operators (introduced in ) and of the relational operators
in Definition \ref{def:relAlg}.
\\
Thus, an IRDB is a member of the NewSQL, that is, a member of a
class of modern relational database management systems that seek to
provide the same scalable performance of NoSQL systems for online
transaction processing (read-write) workloads while still
maintaining the ACID guarantees of a traditional database system.\\
We can easy see that the mapping tgds used from the Big Data vector
table $\overrightarrow{A}$ (the source schema in Data Integration)
into user-defined RDB schema $\A$ (the global schema of this Data
Integration system with integrity constraints) is not simple FOL
formula. Because the same element $r_k$ is used as a predicate
symbol (on the right-side of the tgd's implication) and as a value
(on the left side of the implication as the first value in the
predicate $r_V$). It means that the elements of the domain of this
logic are the elements of other classes and are the classes for
themselves as well. Such semantics is not possible in the standard
FOL, but only in the \emph{intensional} FOL.

\section{Intensional semantics  for IRDBs}

 More about
 relevant recent works for intensional FOL can be found  in
\cite{Majk09FOL,Majk12a} where a new conservative intensional
extension of the
Tarski's semantics of the FOL is defined.\\
 %
%
 Intensional entities are such concepts as
propositions and properties. The term 'intensional' means that they
violate the principle of extensionality; the principle that
extensional equivalence implies identity. All (or most) of these
intensional entities have been classified at one time or another as
kinds of Universals \cite{Beal93}.\\
We consider a non empty domain $~\D = D_{-1} \bigcup D_I$,  where a
subdomain $D_{-1}$ is made of
 particulars (extensional entities), and the rest $D_I = D_0 \bigcup
 D_1 ...\bigcup D_n ...$ is made of
 universals ($D_0$ for propositions (the 0-ary concepts), and  $D_n, n \geq 1,$ for
 n-ary concepts).\\
 The fundamental entities
are \emph{intensional abstracts} or so called, 'that'-clauses. We
assume that they are singular terms; Intensional expressions like
'believe', mean', 'assert', 'know',
 are standard two-place predicates  that take 'that'-clauses as
 arguments. Expressions like 'is necessary', 'is true', and 'is
 possible' are one-place predicates that take 'that'-clauses as
 arguments. For example, in the intensional sentence "it is
 necessary that $\phi$", where $\phi$ is a proposition, the 'that $\phi$' is
 denoted by the $\lessdot \phi \gtrdot$, where $\lessdot \gtrdot$ is the intensional abstraction
 operator which transforms a logic formula into a \emph{term}. Or, for example, "x believes that $\phi$" is given by formula
$p_i(x,\lessdot \phi \gtrdot)$ ( $p_i$ is binary 'believe'
predicate). We introduce an  intensional FOL  \cite{Majk12a}, with
slightly different intensional abstraction than that originally
presented  in \cite{Beal79}, as follows:
 \begin{definition} \label{def:bealer}
  The syntax of the First-order Logic (FOL) language $\L$ with intensional abstraction
$\lessdot \gtrdot$ is as follows:\\
 Logical operators $(\wedge, \neg, \exists)$; Predicate letters $r_i,p_i \in
 \mathbb{R}$
 with a given arity $k_i = ar(r_i) \geq 1$, $i = 1,2,...$ (the functional letters are considered as particular case of the predicate
 letters); a set PR  of propositional letters (nullary predicates) with a truth $r_\emptyset \in PR \bigcap \mathbb{R}$;  Language constants $\overline{0}, \overline{1},...,\overline{c},\overline{d}...$; Variables $x,y,z,..$ in $\V$; Abstraction $\lessdot \_ \gtrdot$, and punctuation
 symbols (comma, parenthesis).
 With the following simultaneous inductive definition of \emph{term} and
 \emph{formula}:\\
   1. All variables and constants  are terms. All propositional letters are formulae.\\
   2. If $~t_1,...,t_k$ are terms then $r_i(t_1,...,t_k)$ is a formula
 for a k-ary predicate letter $r_i \in \mathbb{R}$ .\\
   3. If $\phi$ and $\psi$ are formulae, then $(\phi \wedge \psi)$, $\neg \phi$, and
 $(\exists x)\phi$ are formulae. \\
   4. If $\phi(\textbf{x})$ is a formula (virtual predicate) with a list of free variables in $\textbf{x} =(x_1,...,x_n)$ (with ordering
from-left-to-right of their appearance in $\phi$), and  $\alpha$ is
its sublist of \emph{distinct} variables,
 then $\lessdot \phi \gtrdot_{\alpha}^{\beta}$ is a term, where $\beta$ is the remaining list of free variables preserving ordering in $\textbf{x}$ as well. The externally quantifiable variables are the \emph{free} variables not in $\alpha$.
  When $n =0,~ \lessdot \phi \gtrdot$ is a term which denotes a
proposition, for $n \geq 1$ it denotes
 a n-ary concept.\\
An occurrence of a variable $x_i$ in a formula (or a term) is
\emph{bound} (\emph{free}) iff it lies (does not lie) within a
formula of the form $(\exists x_i)\phi$ (or a term of the form
$\lessdot \phi \gtrdot_{\alpha}^{\beta}$ with $x_i \in \alpha$). A
variable is free (bound) in a formula (or term) iff it has (does not
have) a free occurrence in that formula (or term). A \emph{sentence}
is a formula having no free variables.
\end{definition}
An interpretation (Tarski) \index{Tarski's interpretations} $I_T$
consists of a nonempty domain
   $\D = D_{-1} \bigcup D_I$ and a mapping that assigns to any predicate letter $r_i \in
   \mathbb{R}$ with $k = ar(r_i)\geq 1$, a relation $\|r_i\| = I_T(r_i) \subseteq
   \D^k$;
   to each
   individual constant $\overline{c}$ one given element $I_T(\overline{c}) \in
   \D$, with $I_T(\overline{0}) = 0, I_T(\overline{1}) = 1$ for
   natural numbers $\N =\{0,1,2,...\}$, and to any
   propositional letter $p \in PR$ one  truth value $I_T(p) \in
    \{f,t\}$, where $f$ and $t$ are the empty set $\{\}$
   and the singleton set $\{<>\}$ (with the empty tuple $<> \in
   D_{-1}$), as those used  in the Codd's relational-database algebra \cite{Codd72} respectively,
    so that for any $I_T$, $I_T(r_\emptyset) = \{<>\}$
   (i.e., $r_\emptyset$ is a tautology), while $Truth \in D_0$ denotes the concept (intension)
of this tautology. \\
Note that in the intensional semantics a k-ary functional symbol,
for $k \geq 1$, in standard (extensional) FOL is considered as a
$(k+1)$-ary predicate symbols: let $f_m$ be such a $(k+1)$-ary
predicate symbol which represents a k-ary function denoted by
$\underline{f}_m$ with standard Tarski's interpretation
$I_T(\underline{f}_m):\D^k \rightarrow \D$. Then $I_T(f_m)$ is a
relation obtained from its graph, i.e.,  $I_T(f_m) = R =
\{(d_1,...,d_k,I_T(\underline{f}_m)(d_1,...,d_k)) ~| ~d_i \in \D, 1\leq i \leq k \}$.\\
The universal quantifier is defined as usual by $\forall = \neg
\exists \neg$. Disjunction $\phi \vee \psi$ and implication $\phi
\Rightarrow \psi$ are expressed by
 $\neg(\neg \phi \wedge \neg \psi)$ and $\neg \phi \vee
 \psi$, respectively.
 In FOL with the
identity $\doteq$, the formula $(\exists_1 x)\phi(x)$ denotes the
formula $(\exists x)\phi(x) \wedge (\forall x)(\forall y)(\phi(x)
\wedge \phi(y)  \Rightarrow (x \doteq y))$. We denote by $R_{=}$ the
Tarski's interpretation of
$\doteq$.\\
In what follows any open-sentence, a formula $\phi$ with non empty
tuple  of free variables $(x_1,...,x_m)$, will be called a m-ary
  \emph{virtual predicate}, denoted also by
$\phi(x_1,...,x_m)$. This definition contains the precise method of
establishing the \emph{ordering} of variables in this tuple:
  such an method that will be adopted here is the ordering of appearance, from left to right, of free variables in $\phi$.
   This method of composing the tuple of free variables
  is the unique and canonical way of definition of the virtual predicate from a given
  formula.\\
An \emph{intensional interpretation} of this intensional FOL is a
mapping between the set $\L$ of formulae of the logic language  and
 intensional entities in $\D$, $I:\L \rightarrow \D$, is a kind of
 "conceptualization", such that  an open-sentence (virtual
 predicate)
 $\phi(x_1,...,x_k)$ with a tuple $\textbf{x}$ of all free variables
 $(x_1,...,x_k)$ is mapped into a k-ary \emph{concept}, that is, an intensional entity  $u =
 I(\phi(x_1,...,x_k)) \in D_k$, and (closed) sentence $\psi$ into a proposition (i.e., \emph{logic} concept) $v =
 I(\psi) \in D_0$ with $I(\top) = Truth \in D_0$ for a FOL tautology $\top$.
This interpretation $I$ is extended also to the terms (called as
denotation as well).
 A language constant $\overline{c}$ is mapped into a
 particular (an extensional entity) $a = I(\overline{c}) \in D_{-1}$ if it is a proper name, otherwise in a correspondent concept in
$\D$.
For each $k$-ary  atom $r_i(\textbf{x})$, $I(\lessdot
r_i(\textbf{x}) \gtrdot_{\textbf{x}})$ is the relation-name (symbol)
$r_i \in \mathbb{R}$ (only if $r_i$ is not  defined as a language
constant as well).
The extension of $I$ to  the complex abstracted terms  is given in \cite{Majk12a} (in Definition 4).\\
 An assignment $g:\V \rightarrow \D$ for variables in $\V$ is
applied only to free variables in terms and formulae.  Such an
assignment $g \in \D^{\V}$ can be recursively uniquely extended into
the assignment $g^*:\T X \rightarrow \D$, where $\T X$ denotes the
set of all terms with variables in $X \subseteq \V$ (here $I$ is an
intensional interpretation of this FOL, as explained
in what follows), by :\\
1. $g^*(t_k) = g(x) \in \D$ if the term $t_k$ is a variable $x \in
\V$.\\
2. $g^*(t_k) = I(\overline{c}) \in \D$ if the term $t_k$ is a
constant
$\overline{c}$.\\
3. if $t_k$ is an abstracted term $\lessdot \phi
\gtrdot_{\alpha}^{\beta}$,  then $g^*(\lessdot \phi
\gtrdot_{\alpha}^{\beta}) = I(\phi[\beta /g(\beta)] ) \in D_k, k =
|\alpha|$ (i.e., the number of variables in $\alpha$), where
$g(\beta) = g(y_1,..,y_m) = (g(y_1),...,g(y_m))$ and $[\beta
/g(\beta)]$ is a uniform replacement of each i-th variable in the
list $\beta$
with the i-th constant in the list $g(\beta)$. Notice that $\alpha$ is the list of all free variables in the formula $\phi[\beta /g(\beta)]$.\\
  We denote by $~t_k/g~$ (or $\phi/g$) the ground term (or
formula) without free variables, obtained by assignment $g$ from a
term $t_k$ (or a formula $\phi$), and by  $\phi[x/t_k]$ the formula
obtained by  uniformly replacing $x$ by a term $t_k$ in $\phi$.\\
The distinction between intensions and extensions is important
 especially because we are now able to have an \emph{equational
 theory} over intensional entities (as  $\lessdot \phi \gtrdot$), that
 is predicate and function "names", that is separate from the
 extensional equality of relations and functions.
 An \emph{extensionalization function} $h$ assigns to the intensional elements of $\D$ an appropriate
extension as follows: for each proposition $u \in D_0$, $h(u) \in
 \{f,t\} \subseteq \P(D_{-1})$ is its
 extension (true or false value); for each n-ary
 concept $u \in D_n$, $h(u)$ is a subset of $\D^n$
 (n-th Cartesian product of $\D$); in the case of particulars $u \in
 D_{-1}$, $h(u) = u$.\\
 We define  $\D^0 = \{<>\}$, so that $\{f,t\} = \P(\D^0)$, where $\P$ is the powerset operator.
 Thus we have (we denote the disjoint union by '+'):
 \\$h = (h_{-1}  + \sum_{i\geq 0}h_i):\sum_{i
\geq -1}D_i \longrightarrow D_{-1} +  \sum_{i\geq 0}\P(D^i)$,\\
 where $h_{-1} = id:D_{-1} \rightarrow D_{-1}$
is identity mapping, the mapping $h_0:D_0 \rightarrow \{f,t\}$
assigns the truth values in $ \{f,t\}$ to all propositions, and the
mappings $h_i:D_i \rightarrow \P(D^i)$, $i\geq 1$, assign an
extension to all concepts. Thus, the intensions can be seen as
\emph{names} of abstract or concrete entities, while the extensions
correspond to
various rules that these entities play in different worlds.\\
\textbf{Remark:} (Tarski's constraints) This intensional semantics
has to preserve standard Tarski's semantics of the FOL. That is, for
any formula $\phi \in \L$ with a tuple of free variables
$(x_1,...,x_k)$,  and  $h \in \E$,  the following conservative
 conditions for all  assignments $g,g' \in \D^{\V}$ has to be satisfied: \\
(T)$~~~~~~~h(I(\phi/g)) = t~~$ iff
$~~(g(x_1),...,g(x_k)) \in h(I(\phi))$;\\
and, if $\phi$ is a predicate letter $p$, $k = ar(p) \geq 2$ which
represents a (k-1)-ary
functional symbol $f^{k-1}$ in standard FOL,\\
 (TF)$~~~~~~~h(I(\phi/g)) = h(I(\phi/g')) = t$ and
 $\forall_{1\leq i \leq k-1}(g'(x_i) = g(x_i))~~$ implies $~~g'(x_{k+1})=
 g(x_{k+1})$.\\
$\square$\\
 Thus, intensional  FOL  has a simple Tarski's
first-order semantics, with a decidable
 unification problem, but we need also the actual world mapping
 which maps any intensional entity to its \emph{actual world
 extension}. In what follows we will identify a \emph{possible world} by a
 particular mapping which assigns, in such a possible world,  the extensions to intensional entities.
 This is direct bridge between
 an intensional FOL  and a possible worlds representation
 \cite{Lewi86,Stal84,Mont70,Mont73,Mont74,Majk09FOL}, where the intension (meaning) of a proposition is a
 \emph{function}, from a set of possible worlds $\W$ into the set of
 truth-values.
 Consequently, $\E$ denotes the set of possible
\emph{extensionalization functions} $h$ satisfying the constraint
(T). Each $h \in \E$ may be seen as a \emph{possible world}
(analogously to Montague's intensional semantics for natural
language \cite{Mont70,Mont74}), as it has been demonstrated in
\cite{Majk08in,Majk08ird}, and given by the bijection
$~~~is:\W \simeq \E$.\\
Now we are able to formally define this intensional semantics
\cite{Majk09FOL}:
 \begin{definition} \label{def:intensemant} \textsc{Two-step \textsc{I}ntensional
 \textsc{S}emantics:}\\
Let $~\mathfrak{R} = \bigcup_{k \in \mathbb{N}} \P(\D^k) =
\sum_{k\in \mathbb{N}}\P(D^k)$ be the set of all k-ary relations,
where $k \in \mathbb{N} = \{0,1,2,...\}$. Notice that $\{f,t\} =
\P(\D^0) \in \mathfrak{R}$, that is, the truth values are extensions
in $\mathfrak{R}$. The intensional semantics of the logic language
with the set of formulae $\L$ can be represented by the  mapping
\begin{center}
$~~~ \L ~\rTo^{I} \D ~\Longrightarrow_{w \in \W}~ \mathfrak{R}$,
\end{center}
where $\rTo^{I}$ is a \emph{fixed intensional} interpretation $I:\L
\rightarrow \D$ and $~\Longrightarrow_{w \in \W}~$ is \emph{the set}
of all extensionalization functions $h = is(w):\D \rightarrow
\mathfrak{R}$ in $\E$, where $is:\W \rightarrow \E$ is the mapping
from the set of possible worlds to the set of
 extensionalization functions.\\
 We define the mapping $I_n:\L_{op} \rightarrow
\mathfrak{R}^{\W}$, where $\L_{op}$ is the subset of formulae with
free variables (virtual predicates), such that for any virtual
predicate $\phi(x_1,...,x_k) \in \L_{op}$ the mapping
$I_n(\phi(x_1,...,x_k)):\W \rightarrow \mathfrak{R}$ is the
Montague's meaning (i.e., \emph{intension}) of this virtual
predicate \cite{Lewi86,Stal84,Mont70,Mont73,Mont74}, that is, the
mapping which returns with the extension of this (virtual) predicate
in each possible world $w\in \W$.
\end{definition}
 Another relevant question w.r.t. this two-step
interpretations of an intensional semantics is how in it is managed
the extensional identity relation $\doteq$ (binary predicate of the
identity) of the FOL. Here this extensional identity relation is
mapped into the binary concept $Id = I(\doteq(x,y)) \in D_2$, such
that $(\forall w \in \W)(is(w)(Id) = R_{=})$, where $\doteq(x,y)$
(i.e., $p_1^2(x,y)$) denotes an atom of the FOL of the binary
predicate for identity in FOL, usually written by FOL formula $x
\doteq y$.\\
 Note that here we prefer to distinguish this \emph{formal
symbol} $~ \doteq ~ \in \mathbb{R}$ of the built-in identity binary
predicate letter in the FOL, from the standard mathematical
symbol '$=$' used in all mathematical definitions in this paper.\\
 In what follows we will use the function $f_{<>}:\mathfrak{R}
\rightarrow \mathfrak{R}$, such that for any relation $R \in
\mathfrak{R}$, $f_{<>}(R) = \{<>\}$ if $R \neq \emptyset$;
$\emptyset$ otherwise. Let us define the following set of algebraic
operators for
 relations in $\mathfrak{R}$:
\begin{enumerate}
\item binary operator $~\bowtie_{S}:\mathfrak{R} \times \mathfrak{R} \rightarrow
\mathfrak{R}$,
 such that for any two relations $R_1, R_2 \in
 \mathfrak{R}~$, the
 $~R_1 \bowtie_{S} R_2$ is equal
to the relation obtained by natural join
 of these two relations $~$ \verb"if"
 $S$ is a non empty
set of pairs of joined columns of respective relations (where the
first argument is the column index of the relation $R_1$ while the
second argument is the column index of the joined column of the
relation $R_2$); \verb"otherwise" it is equal to the cartesian
product $R_1\times R_2$.\\ For example, the logic formula
$\phi(x_i,x_j,x_k,x_l,x_m) \wedge \psi (x_l,y_i,x_j,y_j)$ will be
traduced by the algebraic expression $~R_1 \bowtie_{S}R_2$ where
$R_1 \in \P(\D^5), R_2\in \P(\D^4)$ are the extensions for a given
Tarski's interpretation  of the virtual predicate $\phi, \psi$
relatively, so that $S = \{(4,1),(2,3)\}$ and the resulting relation
will have the following ordering of attributes:
$(x_i,x_j,x_k,x_l,x_m,y_i,y_j)$.
\item unary operator $~ \sim:\mathfrak{R} \rightarrow \mathfrak{R}$, such that for any k-ary (with $k \geq 0$)
relation $R \in  \P(\D^{k}) \subset \mathfrak{R}$
 we have that $~ \sim(R) = \D^k \backslash R \in \D^{k}$, where '$\backslash$' is the substraction of relations. For example, the
logic formula $\neg \phi(x_i,x_j,x_k,x_l,x_m)$ will be traduced by
the algebraic expression $~\D^5 \backslash R$ where $R$ is the
extensions for a given Tarski's interpretation  of the virtual
predicate $\phi$.
\item unary operator $~ \pi_{-m}:\mathfrak{R} \rightarrow \mathfrak{R}$, such that for any k-ary (with $k \geq 0$) relation $R \in \P(\D^{k}) \subset \mathfrak{R}$
we have that $~ \pi_{-m} (R)$ is equal to the relation obtained by
elimination of the m-th column of the relation $R~$ \verb"if" $1\leq
m \leq k$ and $k \geq 2$; equal to $~f_{<>}(R)~$ \verb"if" $m = k
=1$; \verb"otherwise" it is equal to $R$. \\For example, the logic
formula $(\exists x_k) \phi(x_i,x_j,x_k,x_l,x_m)$ will be traduced
by the algebraic expression $~\pi_{-3}(R)$ where $R$ is the
extensions for a given Tarski's interpretation  of the virtual
predicate $\phi$ and the resulting relation will have the following
ordering of attributes: $(x_i,x_j,x_l,x_m)$.
\end{enumerate}
Notice that the ordering of attributes of resulting relations
corresponds to the method used for generating the ordering of
variables in the tuples of free variables adopted for virtual
predicates.
\begin{definition}  \label{def:int-algebra} Intensional algebra for the intensional FOL  in Definition \ref{def:bealer} is a structure $~\A_{int}
= ~(\D,  f, t, Id, Truth,  \{conj_{S}\}_{ S \in \P(\mathbb{N}^2)},
neg, \{exists_{n}\}_{n \in \mathbb{N}})$,  $~~$  with
 binary operations  $~~conj_{S}:D_I\times D_I \rightarrow D_I$,
   unary operation  $~~neg:D_I\rightarrow D_I$,  unary
   operations $~~exists_{n}:D_{I}\rightarrow D_I$,  such that for any
extensionalization function $h \in \E$,
and $u \in D_k, v \in D_j$, $k,j \geq 0$,\\
1. $~h(Id) = R_=~$ and $~h(Truth) = \{<>\}$.\\
2. $~h(conj_{S}(u, v)) = h(u) \bowtie_{S}h(v)$, where $\bowtie_{S}$
is the natural join operation defined above and $conj_{S}(u, v) \in
D_m$ where $m = k + j - |S|$
 if for every pair $(i_1,i_2) \in S$ it holds that $1\leq i_1 \leq k$, $1 \leq i_2 \leq j$ (otherwise $conj_{S}(u, v) \in D_{k+j}$).\\
3. $~h(neg(u)) = ~\sim(h(u)) = \D^k \backslash (h(u))$,
 where  $~\sim~$ is the operation
defined above and $neg(u) \in D_k$.\\
 4. $~h(exists_{n}(u)) =
\pi_{-n}(h(u))$, where $\pi_{-n}$ is the operation defined above and
\\ $exists_n(u) \in D_{k-1}$ if $1 \leq n \leq k$ (otherwise
$exists_n$ is the identity function).
\end{definition}
Notice that for $u,v \in D_0$, so that $h(u),h(v) \in \{f,t\}$,
$~h(neg(u)) = ~\sim(h(u)) = \D^0 \backslash (h(u)) = \{<>\}
\backslash (h(u)) \in \{f,t\}$, and $h(conj_\emptyset(u,v) =
h(u)\bowtie_\emptyset h(v)  \in \{f,t\}$.\\
 Intensional interpretation $I:\L \rightarrow \D$ satisfies the
following homomorphic extension:
\begin{enumerate}
  \item The logic formula $\phi(x_i,x_j,x_k,x_l,x_m) \wedge \psi
(x_l,y_i,x_j,y_j)$ will be intensionally interpreted by the concept
$u_1 \in D_7$, obtained by the algebraic expression $~
conj_{S}(u,v)$ where $u = I(\phi(x_i,x_j,x_k,x_l,x_m)) \in D_5, v =
I(\psi (x_l,y_i,x_j,y_j))\in D_4$ are the concepts of the virtual
predicates $\phi, \psi$, relatively, and $S = \{(4,1),(2,3)\}$.
Consequently, we have that for any two formulae $\phi,\psi \in \L$
and a particular  operator $conj_S$ uniquely determined by tuples of
free variables in these two formulae, $I(\phi \wedge \psi) =
conj_{S}(I(\phi),I(\psi))$.
  \item The logic formula $\neg \phi(x_i,x_j,x_k,x_l,x_m)$ will be
intensionally interpreted by the concept $u_1  \in D_5$, obtained by
the algebraic expression $~neg(u)$ where $u =
I(\phi(x_i,\\x_j,x_k,x_l,x_m)) \in D_5$ is the concept of the
virtual predicate $\phi$. Consequently, we have that for any formula
$\phi \in \L$, $~I(\neg \phi) = neg(I(\phi))$.
  \item The logic formula $(\exists x_k) \phi(x_i,x_j,x_k,x_l,x_m)$ will
be intensionally interpreted by the concept $u_1  \in D_4$, obtained
by the algebraic expression $~exists_{3}(u)$ where $u =
I(\phi(x_i,x_j,x_k,x_l,x_m)) \in D_5$ is the concept of the virtual
predicate $\phi$. Consequently, we have that for any formula $\phi
\in \L$ and a particular operator $exists_{n}$ uniquely determined
by the position of the  existentially quantified variable in the
tuple of free variables in $\phi$ (otherwise $n =0$ if this
quantified variable is not a free variable in $\phi$), $~I((\exists
x)\phi) = exists_{n}(I(\phi))$.
\end{enumerate}
We can define the derived intensional disjunctions $disj_S$ in a
standard way as,\\
$disj_S(I(\phi),I(\psi)) \triangleq I(\phi \vee_S \psi) =
I(\neg(\neg\phi \wedge_S \neg\psi)) =
neg(conj_S(neg(I(\phi),neg(I(\psi)))))$.\\
 Once one has found a method for specifying the interpretations of
singular terms of $\L$ (take in consideration the particularity of
abstracted terms), the Tarski-style definitions of truth and
validity for  $\L$ may be given in the customary way.
What is proposed  specifically in \cite{Majk12a} is a method for
characterizing the intensional interpretations of singular terms of
$\L$ in such a way that a given singular abstracted term $\lessdot
\phi \gtrdot_{\alpha}^{\beta}$ will denote an appropriate property,
relation, or proposition, depending on the value of $m =
|\alpha|$.\\
 Notice than if $\beta = \emptyset$ is the empty
list, then $I(\lessdot \phi \gtrdot_{\alpha}^{\beta} ) = I(\phi)$.
Consequently, the denotation of $\lessdot \phi\gtrdot $
 is equal to the meaning of a proposition $\phi$, that is, $~I(\lessdot \phi\gtrdot) =
I(\phi)\in D_0$.  In the case when $\phi$ is an atom
$p_i(x_1,..,x_m)$ then $I (\lessdot
p_i(x_1,..,x_m)\gtrdot_{x_1,..,x_m}) = I(p_i(x_1,..,x_m)) \in D_m$,
while \\$I (\lessdot p_i(x_1,..,x_m)\gtrdot^{x_1,..,x_m}) = union
(\{I(p_i(g(x_1),..,g(x_m)))~|~ g \in \D^{\{x_1,..,x_m\}} \}) \in
D_0$,  with $h(I (\lessdot p_i(x_1,..,x_m)\gtrdot^{x_1,..,x_m})) =
h(I((\exists x_1)...(\exists x_m)p_i(x_1,..,x_m))) \in \{f,t\}$.\\
For example,\\ $h(I(\lessdot p_i(x_1) \wedge \neg p_i(x_1)
\gtrdot^{x_1})) = h(I((\exists x_1)(\lessdot p_i(x_1) \wedge \neg
p_i(x_1)
\gtrdot^{x_1}))) = f$.\\
The interpretation of a more complex abstract $\lessdot \phi
\gtrdot_\alpha^{\beta}$ is defined in terms of the interpretations
of the relevant syntactically simpler expressions, because the
interpretation of more complex formulae is defined in terms of the
interpretation of the relevant syntactically simpler formulae, based
on the intensional algebra above. For example, $I(p_i(x) \wedge
p_k(x)) = conj_{\{(1,1)\}}(I(p_i(x)), I(p_k(x)))$, $I(\neg
\phi) = neg(I(\phi))$, $I(\exists x_i)\phi(x_i,x_j,x_i,x_k) = exists_3(I(\phi))$.\\
Consequently, based on the intensional algebra in Definition
\ref{def:int-algebra} and on intensional interpretations of
abstracted terms, it holds that the interpretation of any formula in
$\L$ (and any abstracted term) will be reduced to an algebraic
expression over interpretations of primitive atoms in $\L$. This
obtained expression is finite for any finite formula (or abstracted
term), and represents the \emph{
meaning} of such finite formula (or abstracted term).\\
Let $\A_{FOL} = (\L, \doteq, \top, \wedge, \neg, \exists)$ be a free
syntax algebra for "First-order logic with identity $\doteq$", with
the set $\L$ of first-order logic formulae,  with $\top$ denoting
the tautology formula (the contradiction formula is denoted by $
\neg \top$), with the set of variables in $\V$ and the domain of
values in $\D$ . \\
Let us define the extensional relational algebra  for the FOL by,\\
$\A_{\mathfrak{R}} = (\mathfrak{R}, R_=, \{<>\}, \{\bowtie_{S}\}_{ S
\in \P(\mathbb{N}^2)}, \sim, \{\pi_{-n}\}_{n \in \mathbb{N}})$,
\\where $ \{<>\} \in \mathfrak{R}$ is the algebraic value
correspondent to the logic truth, and $R_=$ is the binary relation
for extensionally equal elements.
We  use '$=$' for the extensional identity for relations in $\mathfrak{R}$.\\
Then, for any Tarski's interpretation $I_T$ its unique extension to
all formulae $I_T^*:\L \rightarrow \mathfrak{R}$ is also the
homomorphism $I_T^*:\A_{FOL} \rightarrow \A_{\mathfrak{R}}$ from the
free syntax FOL algebra into this extensional relational algebra.\\
 Consequently, we obtain the following Intensional/extensional FOL semantics
 \cite{Majk09FOL}:\\
For any Tarski's interpretation $I_T$ of the FOL, the following
 diagram of homomorphisms commutes,
\begin{diagram}
   && &     \A_{int}~ (concepts/meaning) && &\\
  &\ruTo^{intensional~interpret.~I} && \frac{Frege/Russell}{semantics}  &&\rdTo^{h ~(extensionalization)} &\\
 \A_{FOL}~(syntax)~~~~~~~~      && &\rTo_{I_T^*~(Tarski's ~interpretation)}& && ~~~~~~~~\A_{\mathfrak{R}} ~(denotation)   \\
\end{diagram}
where $h = is(w)$ where $w = I_T \in \W$ is the explicit possible
world (extensional Tarski's interpretation).\\
This homomorphic diagram formally express the fusion of Frege's and
Russell's semantics \cite{Freg92,Russe05,WhRus10} of meaning and
denotation of the FOL language, and renders mathematically correct
the definition of what we call an "intuitive notion of
intensionality", in terms of which a language is intensional if
denotation is distinguished from sense: that is, if both a
denotation and sense is ascribed to its expressions. In fact there
is exactly \emph{one} sense (meaning) of a given logic formula in
$\L$, defined by the uniquely fixed intensional interpretation $I$,
and \emph{a set} of possible denotations (extensions) each
determined by a given Tarski's interpretation of the FOL as follows
from Definition \ref{def:intensemant},
\begin{center}
$~~~ \L ~\rTo^I \D ~\Longrightarrow_{h = is(I_T), I_T \in ~\W }~
\mathfrak{R}$.
\end{center}
Often 'intension' has been used exclusively in connection with
possible worlds semantics, however, here we use (as many others; as
Bealer for example) 'intension' in a more wide sense, that is as an
\emph{algebraic expression} in the intensional algebra of meanings
(concepts) $\A_{int}$ which represents the structural composition of
more complex concepts (meanings) from the given set of atomic
meanings. In fact, in our case, the meaning of the database concept
$u_{DB}$ is expressed  by an algebraic expression in what follows.
Consequently, not only the denotation (extension) is
compositional, but also the meaning (intension) is compositional.\\
%
The application of the intensional FOL semantics to the Data
Integration system $\I = (\A,\S,\M)$ in Definition \ref{def:parsing}
with the user defined RDB schema $\A = (S_A, \Sigma_A)$ and the
vector big table $r_V$ can be summarized in what follows:
\begin{itemize}
  \item Each relational name (symbol) $r_k \in S_A = \{r_1,...,r_n\}$ with the arity $m =
  ar(r_k)$, is an intensional m-ary concept, so that $r_k = I(\lessdot
  r_k(\textbf{x})\gtrdot_{\textbf{x}}) \in D_m$, for a tuple of
  variables $\textbf{x} = \langle x_1,...,x_m \rangle$ and any intensional interpretation $I$.\\
  For a given Tarski's interpretation $I_T$, the extensionalization
  function $h$ is determined by $h(r_k) = \|r_k\| = \{\langle d_1,...,d_m
  \rangle \in \D^m~|~I_T(r_k(d_1,...,d_m)) = t \} = I_T(r_k) \in A$.
  The instance database $A$ of the user-defined RDB schema $\A$ is a
  model of $\A$ if it satisfies all integrity constraints in
  $\Sigma_A$.\\
    \item The relational symbol  $r_V$ of the vector big table
  is a particular (extensional entity), defined also as a language constant, that is, a term for which there exists an
  intensional interpretation with $I(r_V) = r_V \in D_{-1}$,  so that $h(r_V) = r_V$ (the name of the database $\A$).
  We define the intensional concept of the atom $r_V(y_1,...,y_4)$
  of the relational table $r_V$ as $u_{r_V} = I(\lessdot
  r_V(y_1,...,y_4)\gtrdot_{y_1,...,y_4}) \in D_4$, such
  that for a given model
  $A = \{\|r_1\|,...,\|r_n\| \}$ of the user-defined RDB schema $\A$, corresponding to a given Tarski's interpretation $I_T$,
  its extension is determined by $h(u_{r_V}) = I_T(r_V) = \|r_V\| =
  \overrightarrow{A}$.\\
  \item The database unary concept of the user-defined schema $\A$ is  defined by the intensional expression
  $u_{DB} = exists_{2,3,4}(u_{r_V}) \in D_1$, so that its extension
  is equal
  to $h(u_{DB}) = h(exists_{2,3,4}(u_{r_V})) = \pi_1(h(u_{r_V})) =
  \pi_1(\|r_V\|) \subseteq S_A$, that is, to the subset of the nonempty relations in
  the instance database $A$.\\
  \item Intensional nature of the IRDB is evident in the fact that
  each tuple $\langle r_k,Hash(d_1,...,\\d_m),nr_{r_k}(i),d_i
  \rangle \in \overrightarrow{A}$, corresponding to the atom
  $r_V(y_1,y_2,y_3,y_4)/g$
 for an assignment $g$ such that $g(y_1) =
 r_k \in D_m,  g(y_3) =
  nr_{r_k}(i) \in D_{-1}, g(y_2) = Hash(d_1,...,d_m)) \in D_{-1}$ and $g(y_4) = d_i \in \D$, is equal to
  the intensional  tuple $\langle I(\lessdot r_k(\textbf{x}) \gtrdot_{\textbf{x}},Hash(d_1,...,d_m),nr_{r_k}(i),d_i
  \rangle$.\\ Notice that the intensional tuples are different from
  ordinary tuples composed by only particulars (extensional
  elements) in $D_{-1}$, what is the characteristics of the standard
  FOL (where the domain  of values is equal to $D_{-1}$), while here
  the "value" $r_k = I(\lessdot r_k(\textbf{x}) \gtrdot_{\textbf{x}}) \in D_m$
  is an m-ary intensional concept, for which $h(r_k) \neq r_k$ is an m-ary
  relation (while for all ordinary values $d \in D_{-1}$, $h(d) =
  d$).
  \end{itemize}
Based of the intensional interpretation above, we are able to
represent any instance user-defined database $A$ as an intensional
hierarchy system of concepts, presented in the next diagram, where
for each tuple of data $\textbf{d}_i  = (d_{i1},...,d_{im}) \in
D^m_{-1}$, $1\leq i \leq N$, of the relation $h(r_k) = \|r_k\|$, we
have that $h(I(r_V(r_k,Hash(\textbf{d}_i),nr_{r_k}(j),d_{ij}))) =
t$, for
$~d_{ij}$ different from NULL, $j =1,...,m = ar(r_k)$.
\begin{figure}
 \includegraphics[scale=0.92]{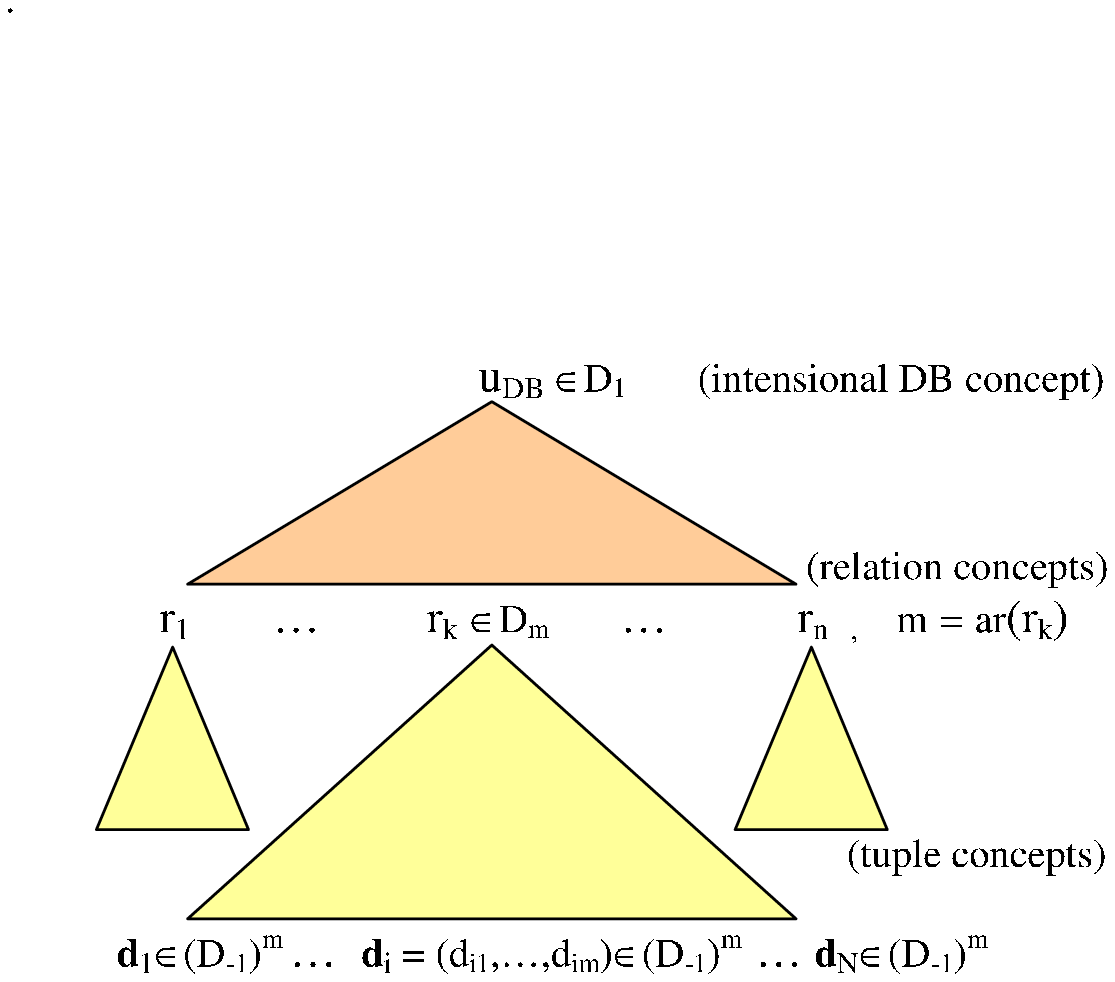}
   \label{fig:treeinf}
  $\vspace*{-24mm}$
 \end{figure}
 \\
The canonical models of such intensional Data Integration system $\I
= \langle \A,\S,\M \rangle$ can be provided in a usual logical
framework as well \cite{Majk14R}:
\begin{propo} \label{prop:canonic}
Let the IRDB be given by a Data Integration system $\I = \langle
\A,\S,\M \rangle$ for a used-defined global  schema $\A =
(S_A,\Sigma_A)$ with $S_A =\{r_1,...,r_n\}$, the source schema $\S =
(\{r_V\},\emptyset)$ with the vector big data relation $r_V$ and the
set of mapping tgds $\M$ from the source schema into  he relations
of the global schema. Then  a canonical model of $\I$ is any model
of the schema $\A^+ = (S_A \bigcup \{r_V\}, \Sigma_A \bigcup \M
\bigcup \M^{OP})$, where $\M^{OP}$ is an opposite mapping tgds  from
$\A$ into $r_V$  given by the following set of tgds:\\
$\M^{OP} = \{\forall x_1,...,x_{ar(r_k)}((r_k(x_1,...,x_{ar(r_k)})
\wedge x_i \emph{NOT NULL}) \Rightarrow\\
r_V(r_k, Hash(x_1,...,x_{ar(r_k)}),nr_{r_k}(i),x_i))~|~1\leq i \leq
ar(r_k), r_k \in S_A \}$.
\end{propo}
The proof can be found in \cite{Majk14R}. The fact that we assumed
$r_V$ to be only a particular (a language constant, i.e., an
extensional entity) is based on the fact that it always will be
materialized (thus non empty relational table) as standard tables in
the RDBs. The other reason is that the extension $h(r_V)$ has not to
be equal to the vector relation (the set of tuples) $\|r_V\|$
because $r_V$ is a name of the database $\A$ composed by the
\emph{set of relations} in the instance database $A$. Consequently,
we do not use the $r_V$ (equal to the name of the database $\A$) as
a value in the tuples of other relations and we do not use the
parsing used for all relations in the user-defined RDB schema $\A$
assumed to be the intensional concepts as well.
 Consequently, the IRDB
has at least one relational table which is not an intensional
concept and which will not be parsed: the vector big table, which
has this singular built-in property in every IRDB.
\section{Reduction of SchemaLog into IRDB}
From the introduction of SchemaLog, we can deduce that the
definition of the Multidatabases has to be obtained mainly by the
following unit clauses (the "facts" in Logic programming) of the following forms \cite{LaSS97}:\\
 (i) $~~(\langle db \rangle::\langle rel
\rangle[\langle tid \rangle: \langle attr \rangle
\rightarrow \langle val \rangle]) \leftarrow $;\\
(ii) $~~(\langle db \rangle::\langle rel
\rangle[\langle attr \rangle]) \leftarrow $;\\
(iii) $~~(\langle db \rangle::\langle rel \rangle) \leftarrow $;\\
(iv) $~~\langle db \rangle \leftarrow$;\\
where the clause (i) correspond to the SQL-like operation of
inserting the value  $\langle val \rangle$ into the attribute
$\langle attr \rangle$ of the relation $\langle rel \rangle$ of the
database $\langle db \rangle$ while other cases correspond to the
DDL-like operations of definitions of the attributes of the
relations, the relations of the databases and the databases. From
the fact that we are interested in the operations over the vector
relations $r_{V_1},...,r_{V_n}$ each one dedicated to a single
database of the given Multidatabase system, the unit clause $\langle
db \rangle \leftarrow $ corresponds to the RDB DDL of the
creation of the vector relation  with the name $r_{V_i} = \langle
db\rangle$ with the four fixed attributes
$\verb"r-name"$,$\verb"t-index"$,$\verb"a-name"$ and $\verb"value"$.
We do not use the clauses (ii) and (iii) for vector relations,
and the only interesting clause is (i). In fact, the clause (i)
corresponds to the SQL statement 'INSERT INTO $r_{V_i}$ VALUES
$(\langle rel \rangle, \langle tid \rangle,\langle attr \rangle,
\langle val \rangle)$'.\\
However, here we can see why SchemaLog cannot be used for real
Multidatabase systems, because each insertion, deletion or update
must be realized  by the updating of the whole Logic program
$P$ which defines the extension of the databases, and then for such a
modified program $P$ to compute its least fixpoint. It is not only a hard
computational process (to rebuild the complete extension of all
databases of a given Multidatabase system by the fixpoint semantics,
but also very complicated task of the concurrent updates of these
databases by different users. This is the common problem and weak
point for almost all AI logic-programming approaches to big databases,
and explains why they can not replace the concurrent RDBMSs
 and  why we intend to translate  the SchemaLog framework
 into the concurrent and Big Data IRDBMSs and then to show that IRDBMs
  can support the interoperability for the Multidatabase systems.\\
\textbf{Remark} (*):  We will consider only the meaningful cases of
the SchemaLog used for Multidatabases, when each relation $r$ of
any database $\A$ is not empty and for each attribute of such a
relation there is at least one value different from NULL, that is,
when every relation and its attributes are really \emph{used} in such a
database to contain the information.\\$\square$\\
Consequently, we consider the
IRDB interoperability with a set of relational databases $S_{DB} =
\{u_{DB_1},...,u_{DB_n}\}$, where each $u_{DB_i} =
exists_{2,3,4}(u_{r_{V_i}}) \in D_1$,  for $i = 1,...,n$, is the intensional DB concept
of the i-th RDB parsed into the vector relation with the name
$r_{V_i}$ (with $u_{r_{V_i}} = I(\lessdot
  r_{V_i}(y_1,...,y_4)\gtrdot_{y_1,...,y_4}) \in D_4$ for a given intensional interpretation $I$).\\
Thus, in this interoperability framework, we will have $n\geq
1$ tree-systems of concepts (provided in previous section) with the
top Multidatabase intensional concept $ u_{mdb} = I(call_1(x))\in D_1$
(where $call_1$ is the unary predicate letter introduced for this
concept introduced for SchemaLog reduction in \cite{LaSS97}) such that $h(u_{mdb}) = S_{DB}$ is the set of database
names in a given Multidatabase system, represented in the
next figure:\\
\begin{figure}
 \includegraphics[scale=0.92]{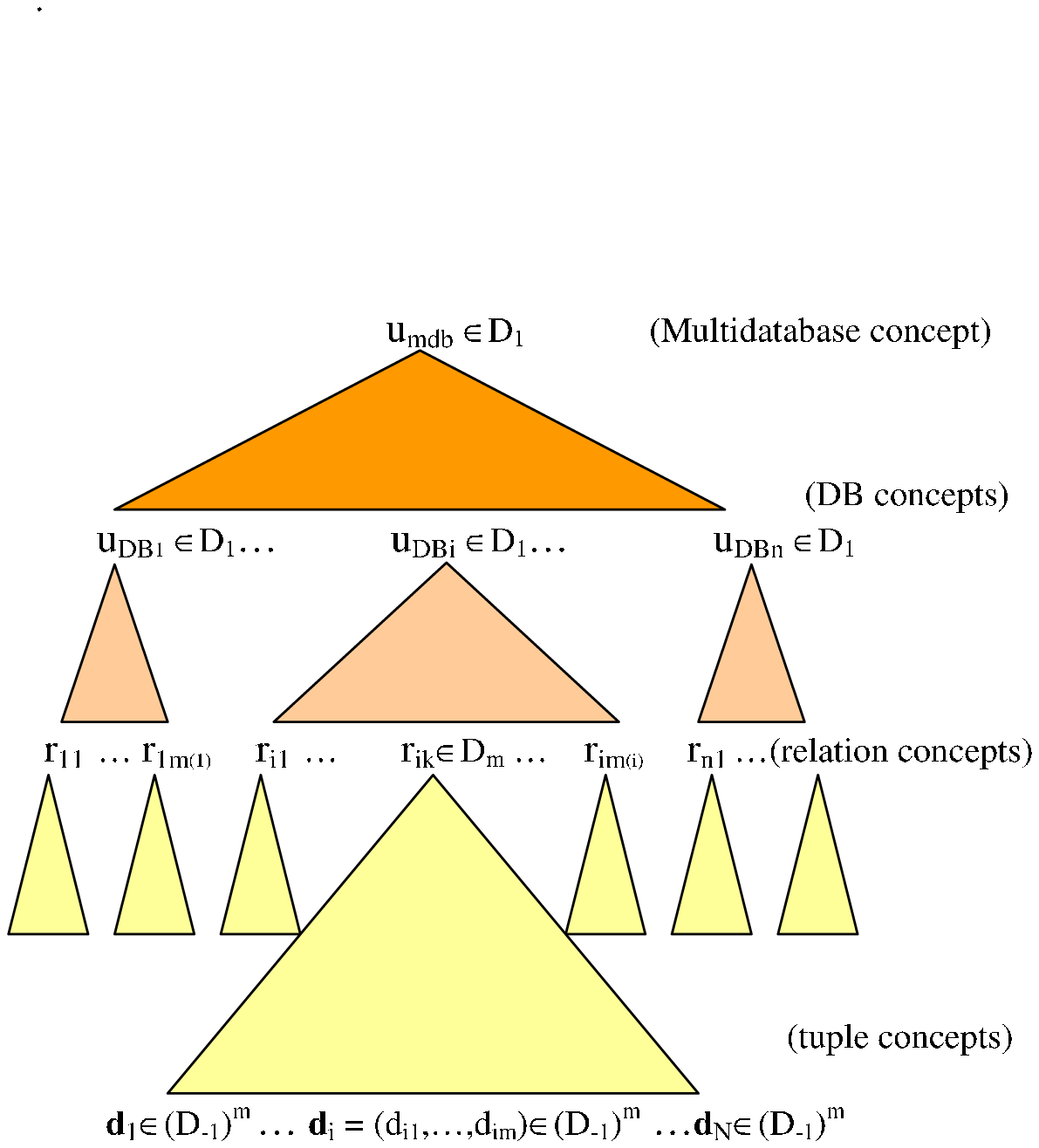}
   \label{fig:treeinf}
  $\vspace*{-24mm}$
 \end{figure}
\\
 Thus, we can
introduce the following intensional concepts (the sorts of
 relations, tuples, attributes and values):\\
1. $ u_{rel} = disj_{S_1}(u_{DB_1},
disj_{S_1}(...,disj_{S_1}(u_{DB_{n-1}},u_{DB_n})...) \in D_1$;\\
2. $ u_{tid} = disj_{S_2}(exists_{1,3,4}(u_{r_{V_1}}),
disj_{S_2}(...,disj_{S_2}(exists_{1,3,4}(u_{r_{V_{n-1}}}),\\exists_{1,3,4}(u_{r_{V_n}}))...) \in D_1$;\\
3. $ u_{attr} = disj_{S_3}(exists_{1,2,4}(u_{r_{V_1}}),
disj_{S_3}(...,disj_{S_3}(exists_{1,2,4}(u_{r_{V_{n-1}}}),\\exists_{1,2,4}(u_{r_{V_n}}))...) \in D_1$;\\
4. $ u_{val} = disj_{S_4}(exists_{1,2,3}(u_{r_{V_1}}),
disj_{S_4}(...,disj_{S_4}(exists_{1,2,3}(u_{r_{V_{n-1}}}),\\exists_{1,2,3}(u_{r_{V_n}}))...) \in D_1$;\\
where $S_i \ \{(i,i)\}$ for $i =1,2,3,4$.\\
Notice that these intensional unary concepts above are derived from
the FOL formulae, as follows:\\
$u_{rel}  = I((\exists x_2,x_3,x_4) r_{V_1}(x_1,x_2,x_3,x_4) \vee
(...\vee (\exists x_2,x_3,x_4)
r_{V_n}(x_1,x_2,x_3,x_4)...))$;\\
$u_{tid}  = I((\exists x_1,x_3,x_4) r_{V_1}(x_1,x_2,x_3,x_4) \vee
(...\vee (\exists x_1,x_3,x_4)
r_{V_n}(x_1,x_2,x_3,x_4)...))$;\\
$u_{attr}  = I((\exists x_1,x_2,x_4) r_{V_1}(x_1,x_2,x_3,x_4) \vee
(...\vee (\exists x_1,x_2,x_4)
r_{V_n}(x_1,x_2,x_3,x_4)...))$;\\
$u_{val}  = I((\exists x_1,x_2,x_3) r_{V_1}(x_1,x_2,x_3,x_4) \vee
(...\vee (\exists x_1,x_2,x_3)
r_{V_n}(x_1,x_2,x_3,x_4)...))$;\\
Then, given a SchemaLog formula $\phi$, its encoding in the
intensional FOL of the IRDB is determined by the recursive
transformation rules given bellow. In this transformation
$\overline{s} \in S \subseteq \T$, $f \in \G$,
$t_i,t_{rel},t_{attr},t_{id},t_{val} \in \T$, $t_{db} \in
\{r_{V_1},...,r_{V_n}\} \subset S \subseteq \T$, are
the SchemaLog terms, and $\phi, \psi$ are any formulae:\\\\
1. $ encode(\overline{s}) = \overline{s}$\\
2. $ encode(f) = f$\\
3. $ encode(f(t_1,...,t_m)) = encode(f)(encode(t_1),...,encode(t_m))$\\
4. $ encode(t_{db}::t_{rel}[t_{id}:t_{attr} \rightarrow t_{val}])
=\\=
encode(t_{db})(encode(t_{rel}),encode(t_{id}),encode(t_{attr}),encode(t_{val}))$\\
5. $encode(t_{db}::t_{rel}[t_{attr}]) =
(\exists x_2,x_4) encode(t_{db})(encode(t_{rel}),x_2,encode(t_{attr}),x_4)$\\
6. $encode(t_{db}::t_{rel}) =
(\exists x_2,x_3,x_4) encode(t_{db})(encode(t_{rel}),x_2,x_3,x_4)$\\
7. $encode(t_{db}) = call_1(t_{db})~~~$ \\
8. $encode(\phi \wedge \psi) = encode(\phi) \wedge encode(\psi)$\\
9. $encode(\phi \vee \psi) = encode(\phi) \vee encode(\psi)$\\
10. $encode(\neg \phi) = \neg encode(\phi)$\\
11. $encode( \rightarrow \phi) = encode(\phi)$\\
12. $encode( \psi \rightarrow \phi) = \neg encode(\psi) \vee (encode(\psi) \wedge encode(\phi))$\\
13. $encode((Qx)\phi) = (Qx)encode(\phi)$, where $Q \in \{\exists,
\forall \}$.\\
In the case of the intensional FOL defined in Definition
\ref{def:bealer}, without  Bealer's intensional abstraction operator
$\lessdot \gtrdot$, we obtain the syntax of the standard FOL but
with intensional semantics as presented in \cite{Majk09FOL}.
Such a FOL has a well known Tarski's interpretation, defined as
follows:
\begin{itemize}
  \item  An interpretation (Tarski) $I_T$ consists in a non empty
domain
   $\D$ and a mapping that assigns to any k-ary predicate letter $p_i$ a relation $R = I_T(p_i) \subseteq \D^k$, to any k-ary functional
   letter $f_i$ a function $I_T(f_i): \D^k \rightarrow \D$, or, equivalently,
its graph relation $R = I_T(f_i)\subseteq \D^{k+1}$ where the
$k+1$-th column is
   the resulting function's value, and to each
   individual constant $\overline{c}$ one given element $I_T(\overline{c}) \in
   \D$.\\
   Consequently, from the intensional point of view, an
   interpretation of Tarski is a possible world in the Montague's
   intensional semantics, that is $w = I_T \in \W$.
   The corespondent extensionalization function is $h = is(w) = is(I_T)$.\\
\item For a given interpretation $I_T$, we define  the satisfaction $I_T^* \models_g$ of a logic formulae in $\L$ for a
given assignment $g:\V \rightarrow \D$ inductively, as follows:\\
 If a formula $\phi$ is an atomic formula $p_i(t_1,...,t_k)$,
then this assignment $g$ satisfies $\phi$, denoted by $I_T^*
\models_g \phi$, iff $(g^*(t_1),...,g^*(t_k)) \in I_T(p_i)$;
 $~g$ satisfies $\neg \phi~$ iff it does not satisfy $\phi$;
 $~g$ satisfies $\phi \wedge \psi~$ iff $g$ satisfies $\phi$ and
$g$ satisfies $\psi$;  $~g$ satisfies $(\exists x_i)\phi~$ iff
exists an assignment $g' \in \D^{\V}$ that may differ from $g$ only
for the variable $x_i \in
\V$, and $g'$ satisfies $\phi$.\\
A formula $\phi$ is \verb"true" for a given interpretation $I_T~$
iff $~\phi$ is satisfied by every assignment $g \in \D^{\V}$. A
formula $\phi$ is \verb"valid" (i.e., tautology) iff $~\phi$ is true
for every Tarksi's interpretation $I_T$.  An interpretation $I_T$ is
a \verb"model" of a set of formulae $\Gamma~$ iff every formula
$\phi \in \Gamma$ is true in this interpretation.
\end{itemize}
\textbf{Semantics}: Given a SchemaLog structure $M = \langle
\D,\I,\I_{fun},\F \rangle$ we construct a corresponding Tarski's
interpretation $I_T = encode(M)$
on the domain $\D$ as follows:\\
$I_T(\overline{s}) \triangleq \I(\overline{s})$, for each $\overline{s} \in \S$; \\
$I_T(f(d_1,...,d_k)) \triangleq \I_{fun}(f)(d_1,...,d_k)$, for each
k-ary functional symbol  $f \in \G$ and $d_1,...,d_k \in
\D$;\\
Note that the $Hash$ functional symbol has to be inserted
into $\G$, so that the built-in function  on strings  $\I_{fun}(Hash)$ satisfies the condition:\\
If  $(\F(r_{V_i})(r)(id)(nr_r(1)) = v_1) \wedge ...\wedge
(\F(r_{V_i})(r)(id)(nr_r(ar(r))) = v_{ar(r)})$ for $r_{V_i}, r,
id,nr_r(k),v_k \in \D$, for $k = 1,...,ar(r)$, then $id =
\I_{fun}(Hash)(v_1,...,v_{ar(r)})$,\\
where some of  $v_i$ can be equal to the value NULL $\in D_{-1}$.\\
We recall that, for intensional FOL, each $k$-ary functional symbol
$f$ is considered as a $(k+1)$-ary relational concept, so that $I(f)
\in D_{k+1}$ with \\$I_T(f) = h(I(f)) = \{\langle
d_1,...,d_k,\I_{fun}(f)(d_1,...,d_k) \rangle
\in h(I(f))| d_1,...,d_k \in \D\}$.\\
 The unique relations that
are materialized in IRDBs are the vector relations, so we will
consider only the relations $r_{V_1},...,r_{V_n}$ (corresponding to
databases $\A_1,...,\A_n$ of this Multidatabase interoperability
system), so that the Tarski's
interpretation for them is constructed in the following way:
\begin{enumerate}
  \item Let $r_{V_i}, r, id,a,v \in  \D$, then\\
$\langle r, id,a,v \rangle \in I_T(r_{V_i})~$ iff
$~\F(r_{V_i})(r)(a)(id)$ is defined in $M$ and
$\F(r_{V_i})(r)(a)(id) =
v$.
  \item For the unary predicate $call_1$, such that from
the Tarski's constraints $u_{mdb} = I(call_1(x))$, we have that $I_T(call_1) = h(I(call_1(x))) =
h(u_{mdb}) = S_{DB}$ (the set of intensional DB concepts in the
figure above). Then,\\
$r_{V_i} \in I_T(call_1)~$ iff $~\F(r_{V_i})$ is defined in $M$.
\end{enumerate}
\begin{propo} \label{prop:encoding}
Let $\phi$ be a SchemaLog formula, $M$ be a SchemaLog structure, and
$g \in \D^{\V}$ an assignment. Let $encode(\phi)$ be the first-order
formula corresponding to $\phi$ and $I_T = encode(M)$ the
corresponding Tarski's interpretation.\\
Then, $M \models_g \phi~$ iff $~I_T^*\models_g encode(\phi)$.
\end{propo}
Proof: Let us show that it holds for all atoms of
SchemaLog:\\
1. Case when $\phi$ is equal to an atom $(t_1::t_2[t_4: t_3 \rightarrow
t_5])$.  Then,\\
$M \models_g (t_1::t_2[t_4: t_3 \rightarrow
t_5])~$ \\
iff $~\F(g(t_1))(g(t_2))(g(t_3))(g(t_4))$ is defined in $M$ and
$\F(g(t_1))(g(t_2))(g(t_3))(g(t_4)) =
g(t_5)$\\
iff $~\langle g(t_2), g(t_4),g(t_3),g(t_5) \rangle \in I_T(g(t_2))$\\
iff $~I_T^*\models_g (t_1/g)(t_2, t_4,t_3,t_5)$\\
iff $~I_T^*\models_g (encode(t_1)/g)(encode(t_2), encode(t_4),encode(t_3),encode(t_5))$\\
iff $~I_T^*\models_g encode(t_1::t_2[t_4: t_2 \rightarrow t_5])$.\\
2. Case when $\phi$ is equal to an atom $(t_1::t_2[t_3])$.  Then,\\
$M \models_g (t_1::t_2[t_3])~$ \\
iff $~\F(g(t_1))(g(t_2))(g(t_3))$ is defined in $M$\\
iff $~I_T^*\models_g (\exists x_2,x_4)(t_1/g)(t_2, x_2,t_3,x_4)$\\
iff $~I_T^*\models_g (\exists x_2,x_4)(encode(t_1)/g)(encode(t_2), x_2,encode(t_3),x_4)$\\
iff $~I_T^*\models_g encode(t_1::t_2[t_3])$.\\
3. Case when $\phi$ is equal to an atom $(t_1::t_2)$.  Then,\\
$M \models_g (t_1::t_2)~$ \\
iff $~\F(g(t_1))(g(t_2))$ is defined in $M$\\
iff $~I_T^*\models_g (\exists x_2,x_3,x_4)(t_1/g)(t_2, x_2,x_3,x_4)$\\
iff $~I_T^*\models_g (\exists x_2,x_3,x_4)(encode(t_1)/g)(encode(t_2), x_2,x_3,x_4)$\\
iff $~I_T^*\models_g encode(t_1::t_2)$.\\
3. Case when $\phi$ is equal to an atom $t_{db}$.  Then,\\
$M \models_g t_{db}~$
iff $~\F(g(t_{db}))$ is defined in $M$\\
 iff $~g(t_{db}) \in I_T(call_1)~$ iff  $~I_T^*\models_g call_1(t_{db})~$ iff $~I_T^*\models_g encode(t_{db})$.\\
4. For the composed formulae, we can demonstrate by induction. Let us
suppose that this property holds for $\phi$ and for $\psi$. Then\\
$M \models_g \phi \vee \psi~$ \\ iff $~M \models_g \phi$ or $~M
\models_g \phi$\\
iff $~I_T^* \models_g encode(\phi)$ or $~I_T^* \models_g encode(\psi)$\\
iff $~I_T^* \models_g encode(\phi \vee \psi)$,\\
and analogously for all other cases.
\\$\square$\\
Note that w.r.t. the Remark (*) above, the relations (predicates)
$call_1, call_2, call_3$ and $call_4$  (obtained by
a similar reduction of SchemaLog in FO Logic Programs in \cite{LaSS97}), can be
defined by the vector relations in IRDBs (see \cite{Majk14R} for the
syntax-semantics if the  relational algebra operators used in next
expressions) as follows:\\
$call_4 =$ (EXTEND $r_{V_1}$ ADD a,\verb"db-name", $r_{V_1}$) UNION
(... UNION (EXTEND $r_{V_n}$ ADD a,\verb"db-name", $r_{V_n}$)...), where $a$ is the attribute used for the database names;\\
$call_3 = call_4 [\verb"db-name",\verb"r-name",\verb"a-name"]$;\\
$call_2 = call_3 [\verb"db-name",\verb"r-name"]$;\\
(note that we also have $call_1 = call_2 [\verb"db-name"]$), with
$h(u_{rel}) = \|call_4 [\verb"r-name"]\|_{\#}$, $h(u_{attr}) = \|call_4 [\verb"a-name"]\|_{\#}$, and $h(u_{val}) = \|call_4 [\verb"value"]\|_{\#}$.\\
Thus, based on \cite{LaSS97}, we obtain the result that
\emph{technically} SchemaLog has no more expressive power than the
intensional first-order logic used for IRDBs.\\
However, differently from the SchemaLog that needs a particular
extension $\mathcal{ERA}$ of the conventional (standard) relational algebra with the new
operations, $\delta, \rho, \alpha$ and $\gamma$  \cite{LaSS97} (so that the resulting algebra is capable of accessing the
database names, relational names and attribute names besides the
values in a federation of database), here we can use the conventional
(standard) SQL over the vector relations $r_{V_1},...,r_{V_n}$ and $call_1$.\\
These new operators $\beta, \rho, \alpha$ are defined in IRDBs by the  following SQL expressions:\\
$\delta() = \|call_1\|$;\\
$\rho(S) = call_2$ WHERE $nr_{call_2}(1)$ IN $S$,  for each $S \subseteq \|call_1\|_{\#}$;\\
$\alpha(S) = call_3$ WHERE $(nr_{call_2}(1), nr_{call_2}(2))$ IN $S$,  for each $S \subseteq \|call_2\|_{\#}$;\\
Only the operation $\gamma$ is a more complex,  defined in \cite{LaSS97} as follows:\\
A \emph{pattern} is a sequence $(p_1,...,p_k), k\geq 0$, where each $p_i$ is one of the forms
$'a_i \rightarrow v_i'$, $'a_i \rightarrow ~'$, $'~ \rightarrow v_i'$, $'~ \rightarrow ~'$. Here $a_i$ is called the
\emph{attribute component} and $v_i$ is called the \emph{value component} of $p_i$. Let $r$ be any relation name, then\\
$'a_i \rightarrow v_i'$ is satisfied by a tuple $tid$ in relation $r$ if $tid[a_i] = v_i$;\\
$'a_i \rightarrow ~'$ is satisfied by a tuple $tid$ in relation $r$ if $a_i$ is an attribute name in $r$;\\
$'~ \rightarrow v_i'$ is satisfied by a tuple $tid$ in relation $r$ if there exists an
attribute $a_i$ in the scheme of $r$ such that $t[a_i] = v_i$;\\
$'~ \rightarrow ~'$ is trivially satisfied by every tuple $tid$ in relation $r$.\\
A pattern $(p_1,...,p_k)$ is satisfied by a tuple $tid$ in relation $r$ if every $p_i, i = 1,...,k$, is satisfied by $tid$.\\
Operator $\gamma$ takes a binary relation $S$ as input, and a pattern as a parameter and returns a relation that consists of tuples
corresponding to those parts of the database where the queried pattern is satisfied. That is,\\
Let $S$ be a binary relation and $(p_1,...,p_k)$ be a pattern, then \cite{LaSS97},\\\\
$\gamma_{(p_1,...,p_k)}(S) \triangleq \{d,r,a_1,v_1,...,a_k,v_k~|~\langle d, r\rangle \in S$ and $d$ is a database in the federation, and $r$ is
a relation in $d$, and $a_i$'s are attributes in $r$, and there exists a tuple $tid$ in $r$ such that
$tid[a_1] = v_1,...,tid[a_k]= v_k$, and $tid$ satisfies $(p_1,...,p_k)\}$.\\\\
Note that when the pattern is empty ($k = 0$), $\gamma_{()}(S)$ would return the set of all pairs
$\langle d, r\rangle \in S$ such that $r$ is a nonempty relation in the database $d$ in the federation.
\begin{theorem} \label{th:ERA-SQL} All new relational operators introduced in SchemaLog extended relational algebra $\mathcal{ERA}$  can be equivalently
expressed by standard SQL terms in IRDBs.
\end{theorem}
\textbf{Proof}: The $call_1$ and $call_2$ are SQL terms (over the vector relations $r_{V_i}$ of the federated databases) defined previously, so that the
definition of the SchemaLog operators $\delta, \rho$ and $\alpha$, given above, are the
standard SQL  terms as well. It is enough to demonstrate that each $\gamma_{(p_1,...,p_k)}$ operator
defined above, can be equivalently represented by a standard SQL term in the IRDBs as follows:\\
(i) Case when $k = 0$. Then $\gamma_{()}(S) = \rho(S)$ (because from Remark (*) we are dealing with the databases with all nonempty relations);\\
(ii) Case when $k = 1$. Then for the SQL term $t = call_4$\\
$\gamma_{(p_1)}(S) = (t$ WHERE $C_{p_1})[\verb"db-name",\verb"r-name",\verb"a-name",\verb"value"]$, \\ where the condition  $C_{p_1}$ is defined by
(here $\top$ is a tautology, for example $\overline{1} = \overline{1}$):\\
\begin{displaymath}
   C_{p_1} = \left\{ \begin{array}{ll}
   (nr_t(4) = a_i) \wedge (nr_t(5) = v_i) & \textrm{ ~, ~if ~$ p_1 = 'a_i \rightarrow v_i'$}\\
  nr_t(4) = a_i & \textrm{~,~ if ~$p_1 = 'a_i \rightarrow ~'$}\\
  nr_t(5) = v_i & \textrm{~,~ if ~$p_1 = '~ \rightarrow v_i'$}\\
\top & \textrm{~,~otherwise}
\end{array} \right.
\end{displaymath}
(iii) Case when $k \geq 2$. Let us define the SQL Cartesian product
 $t = \overbrace{call_4 \bigotimes ... \bigotimes call_4}^k$. Then\\
$\gamma_{(p_1,...,p_k)}(S) = (t$ WHERE $(nr_t(1) = ...=nr_t(5k-4)) \wedge\\
(nr_t(2) = ...=nr_t(5k-3)) \wedge (nr_t(3) = ...=nr_t(5k-2)) \wedge\\
(C_{p_1}\wedge...\wedge C_{p_k}))[nr_t(1),nr_t(2),nr_t(4),nr_t(5),nr_t(9),nr_t(10),...,nr_t(5k-1),nr_t(5k)]$, \\
 where the conditions  $C_{p_m}$, for $~m =1,...,k$, are defined by:
 \begin{displaymath}
   C_{p_m} = \left\{ \begin{array}{ll}
   (nr_t(5m-1) = a_i) \wedge (nr_t(5m) = v_i) & \textrm{ ~, ~if ~$ p_m = 'a_i \rightarrow v_i'$}\\
  nr_t(5m-1) = a_i & \textrm{~,~ if ~$p_m = 'a_i \rightarrow ~'$}\\
  nr_t(5m) = v_i & \textrm{~,~ if ~$p_m = '~ \rightarrow v_i'$}\\
\top & \textrm{~,~otherwise}
\end{array} \right.
\end{displaymath}
$\square$\\
\begin{example} \label{exam:gamma} Let us consider the Multidatabase (federated) system given in Example 2.1 in
\cite{LaSS97}, consisting of RDB \verb"univ_A",\verb"univ_B" and \verb"univ_C"
corresponding to universities A,B and C. Each database maintains information on the university's departments, staff, and the
average salary in 1997, as follows:\\
1. The RDB \verb"univ_A" has the following single relation \verb"pay-info" which has one
tuple for each department and each category in that department:\\\\
$\|\verb"pay-info"\| = $\begin{tabular}{|lll|}
  \hline
  \verb"category" & ~\verb"dept" & ~\verb"avg-sal" \\
  \hline
  Prof & ~CS & ~70,000 \\
  Assoc. Prof & ~CS & ~60,000 \\
  Secretary & ~CS & ~35,000 \\
  Prof & ~Math & ~65,000 \\
  \hline
\end{tabular}\\\\
2. The RDB \verb"univ_B" has the  single relation, (also \verb"pay-info" ), but in this case,
 department names appear as attribute names and the values corresponding to them are
 the average salaries:\\\\
$\|\verb"pay-info"\| = $\begin{tabular}{|lll|}
  \hline
  \verb"category" & ~\verb"CS" & ~~\verb"Math" \\
  \hline
  Prof & ~80,000 & ~~65,000 \\
  Assoc. Prof & ~65,000 & ~~55,000 \\
  Assist. Prof & ~45,000 & ~~42,000 \\
  \hline
\end{tabular}\\\\

3. The RDB \verb"univ_C" has as many relations as there are departments, and has tuples
corresponding to each category and its average salary in each of the $\verb"detp"_i$ relations:\\\\
$\|\verb"CS"\| = $ \begin{tabular}{|ll|}
  \hline
  \verb"category" & ~\verb"avg-sal"  \\
  \hline
  Prof & ~65,000 \\
  Assist. Prof & ~40,000 \\
  \hline
\end{tabular}
$~~~~~~\|\verb"ece"\| = $ \begin{tabular}{|ll|}
  \hline
  \verb"category" & ~\verb"avg-sal"  \\
  \hline
  Secretary & ~30,000 \\
   Prof & ~70,000 \\
  \hline
\end{tabular}\\\\
By parsing of these three RDBs, we obtain the three vector relations $r_{V_1} = \verb"univ_A"$,  $r_{V_2} = \verb"univ_B"$ and
$r_{V_1} = \verb"univ_C"$.\\ Let us consider the tuple $ID_1 = Hash( Secretary  ~CS  ~35,000)$ of the database
$\verb"univ_A"$, so that\\\\
$\|r_{V_1}\| = \|\verb"univ_A"\| \supset $ \begin{tabular}{|llll|}
                                       \hline
                                        \verb"r-name" &~\verb"t-index" & ~\verb"a-name" & ~\verb"value" \\
                                        \hline
                                       \verb"pay-info" & ~$ID_1$ & ~\verb"category" & ~Secretary \\
                                       \verb"pay-info" & ~$ID_1$ & ~\verb"dept" & ~CS \\
                                       \verb"pay-info" & ~$ID_1$ & ~\verb"avg-sal" & ~35,000 \\
                                       \hline
                                     \end{tabular},\\\\
and consider the tuple $ID_2 = Hash(Secretary 30,000)$ of the relation \verb"ece"
of the database $\verb"univ_C"$, so that\\\\
$\|r_{V_3}\| = \|\verb"univ_C"\| \supset $ \begin{tabular}{|llll|}
                                       \hline
                                        \verb"r-name" &~\verb"t-index" & ~\verb"a-name" & ~\verb"value" \\
                                        \hline
                                       \verb"ece" & ~$ID_2$ & ~\verb"category" & ~Secretary \\
                                       \verb"ece" & ~$ID_2$ & ~\verb"avg-sal" & ~30,000 \\
                                       \hline
                                     \end{tabular}\\\\
so that the following set of tuples are the part of the relation obtained from the SQL algebra
term $call_4$:\\
$\|call_4\|_{\#}   \supset $ \begin{tabular}{|lllll|}
                                       \hline
                                      \verb"db-name" &  ~\verb"r-name" &~\verb"t-index" & ~\verb"a-name" & ~\verb"value" \\
                                        \hline
                                       \verb"univ_A"&~\verb"pay-info" & ~$ID_1$ & ~\verb"category" & ~Secretary \\
                                       \verb"univ_A"&~\verb"pay-info" & ~$ID_1$ & ~\verb"dept" & ~CS \\
                                       \verb"univ_A"&~\verb"pay-info" & ~$ID_1$ & ~\verb"avg-sal" & ~35,000 \\
                                       \verb"univ_C"&~\verb"ece" & ~$ID_2$ & ~\verb"category" & ~Secretary \\
                                       \verb"univ_C"&~\verb"ece" & ~$ID_2$ & ~\verb"avg-sal" & ~30,000 \\
                                       \hline
                                     \end{tabular},\\\\

Then the operation $\gamma_{(~ \rightarrow Secretary, ~\rightarrow ~)}(S)$ against the university
databases above is equivalent to the SQL term (for $t = call_4 \bigotimes call_4$)\\
$(t$ WHERE $(nr_t(1) =nr_t(6)) \wedge
(nr_t(2) = nr_t(7)) \wedge (nr_t(3)=nr_t(7)) \wedge\\
(nr_t(5) = Secretary))[nr_t(1),nr_t(2),nr_t(4),nr_t(5),nr_t(9),nr_t(10)]$,\\
will yield for \\
$S =$ \begin{tabular}{|l|l|}
        \hline
        \verb"univ_A"&~\verb"pay-info" \\
        \verb"univ_B"&~\verb"pay-info" \\
        \verb"univ_C"&~\verb"CS" \\
        \verb"univ_C"&~\verb"ece" \\
        \hline
      \end{tabular} $= \| call_2\|_{\#} = \|call_4 [\verb"db-name", \verb"r-name"]\|_{\#}$\\
the relation\\
\begin{tabular}{|l|l|l|l|l|l|}
  \hline
  \verb"univ_A"&~\verb"pay-info" &~\verb"category" &~Secretary &~\verb"dept"&~CS\\
  \verb"univ_A"&~\verb"pay-info" &~\verb"category" &~Secretary &~\verb"category"&~Secretary\\
  \verb"univ_A"&~\verb"pay-info" &~\verb"category" &~Secretary &~\verb"avg-sal"&~35,000\\
  \verb"univ_C"&~\verb"ece" &~\verb"category" &~Secretary &~\verb"category"&~Secretary\\
  \verb"univ_C"&~\verb"ece" &~\verb"category" &~Secretary &~\verb"avg-sal"&~30,000\\
   \hline
\end{tabular}
\end{example}
$\square$\\
Thus, we obtain he following completeness result for the SQL in the IRDBs w.r.t. the
Querying Fragment ($\L_Q$) of SchemaLog (provided in Definition 6.6 in \cite{LaSS97}):
\begin{coro} Let $\mathcal{DB}$ be a relational Multidatabase system with nonempty relations and attributes (Remark (*)), $\P$ be a set of
safe rules in the Querying Fragment $\L_Q$ of SchemaLog and $p$ any (virtual) predicate  defined by $\P$.
Then there exists a standard SQL expression $t$ such that the computed relation $\|t\|_{\#}$ in the IRDB obtained by
parsing of this Multidatabase system is equal to the relation corresponding to
$p$ computed by SchemaLog.
\end{coro}
\textbf{Proof}: From Lemma 6.1 in \cite{LaSS97} for such a query $p \in \L_Q$
there is an expression $E$ of the extended relational algebra $\mathcal{ERA}$,
such that the relation corresponding to $p$ is equal to the relation obtained
by computing the relational expression $E$. From Theorem \label{th:ERA-SQL}, we are able
to translate this expression $E \in ERA$ into an equivalent  standard SQL term $t$ whose
extension $\|t\|_{\#}$ in the IRDB obtained by
parsing of this Multidatabase system is equal to the relation corresponding to
$p$ computed by SchemaLog.
\\$\square$\\
Consequently, any querying of data and metadata information (of nonempty relations and nonempty attributes, as explained in Remark (*)) of the federated relational
database system $\mathcal{DB}$ provided by the interoperability framework of the SchemaLog
can be done in the IRDBs framework by the standard SQL.\\
\textbf{Remark}(**): If we need to use the interoperability framework also for the empty database schemas or empty relations,
in that case we need to create the relation table  $call_3$ not by deriving it as a particular projections from
$call_4$ (SQL term) but directly from the RDB dictionary of the Multidatabase system.\\$\square$\\
Consequently, by permitting the SQL querying over the vector relations in the IRBDs we can
obtain the answers  (see \cite{LaSS97} for more useful cases) like, for example,:\\
$(Q_4)$ "\emph{Find the names of all the relations in which the token 'John' appears}";\\
$(Q_5)$ "\emph{Given two relations \verb"r" and \verb"s" (in database \verb"db"),
whose schemas are unknown, compute their natural join}";\\
etc.

\section{Conclusion}
The method of parsing of a relational instance-database $A$ with the
user-defined schema $\A$ into a vector relation
$\overrightarrow{A}$, used in order to represent the information in
a standard and simple key/value form, today in various applications
of Big Data, introduces the intensional concepts for the
user-defined relations of the schema $\A$.
Moreover, we can consider the vector relations as the concept of \emph{mediator}, proposed by Wiederhold \cite{Wied92}, as
means for integrating data from also non-relational heterogeneous sources.
The expressive power of IRDB which includes the expressive power of SchemaLog
and its ability to resolve data/meta-data conflicts suggests that it has the potential for being used
in the interoperability frameworks for the Multidatabase systems and as a platform
for developing mediators.
This new family of IRDBs extends the traditional RDBS with new
features. However, it is compatible in the way how to present the
data by user-defined database schemas (as in RDBs) and with SQL for
management of such a relational data. The  structure of RDB is
parsed into a vector key/value relation so that we obtain a column
representation of data used in Big Data applications, covering the
key/value and column-based Big Data applications as well, into a
unifying RDB framework. The standard SQL syntax of IRDB makes it possible to express
powerful queries and programs in the context of component database interoperability. We are able to treat the
data in database, the schema of the individual databases in a Multidatabase (a federtion) system, as well
as the databases and relations themselves as first class citizens, without using
higher-order syntax or semantics.\\
Note that the method of parsing is well suited for the migration
from  all existent RDB applications where the data is stored in the
relational tables, so that this solution gives the possibility to
pass easily from the actual RDBs into the new machine engines for
the IRDB. We preserve all metadata (RDB schema definitions) without
modification and only dematerialize the relational tables, of a given database $\A_i$,
by transferring their stored data into  the vector relation $r_{V_i}$
(possibly in a number of disjoint partitions over a number of
nodes). From the fact that we are using the query rewriting IDBMS,
the current user's (legacy) applications does not need any
modification and they continue to "see" the same user-defined RDB
schema as before. Consequently, this IRDB solution is adequate for a
massive migration from the already obsolete and slow RDBMSs into a
new family of fast, NewSQL schema-flexible (with also 'Open
schemas') and Big Data scalable IRDBMSs.


\bibliographystyle{IEEEbib}
\bibliography{mydb}



%
\end{document}